%% file: localsimilarity.tex
\documentclass[10pt,journal,compsoc]{IEEEtran}
\ifCLASSOPTIONcompsoc
  \usepackage[nocompress]{cite}
\else
  \usepackage{cite}
\fi
\ifCLASSINFOpdf
\else
\fi
\usepackage{cite}

\usepackage{color}
\usepackage[dvipsnames]{xcolor}
\usepackage{graphicx}
\usepackage{subfigure}
\usepackage{xspace}
\usepackage{verbatim}
\usepackage{amsmath}
\usepackage{amssymb}
\usepackage{url}
\usepackage{amsthm}
\usepackage{soul}
\usepackage[ruled,vlined,linesnumbered]{algorithm2e}
\usepackage{tikz}
\usepackage{pgfplots}
\usepackage{tikzscale}
\usetikzlibrary{calc,positioning}
\usepackage{lipsum,adjustbox}

\pgfplotsset{compat=1.12}
\pgfplotsset{every axis legend/.append style={at={(0.5,1.03)},anchor=south},}

\newcommand{\myparagraph}[1]{\noindent\textbf{\\
#1}}
\newcommand{\MTED}{\textsc{mted}\xspace}
\newcommand{\LMTED}{\textsc{lmted}\xspace}
\newcommand{\RMS}{\textsc{rms}\xspace}

\newcommand{\rulesep}{\unskip\ \vrule\ }
\begin{document}
%
\title{Comparative Analysis of Merge Trees using Local Tree Edit Distance}
\author{Raghavendra~Sridharamurthy,~\IEEEmembership{Student Member,~IEEE},
        and~Vijay Natarajan,~\IEEEmembership{Member,~IEEE}
\IEEEcompsocitemizethanks{\IEEEcompsocthanksitem R. Sridharamurthy and V. Natarajan are with the Department of Computer Science and Automation, Indian Institute of Science, Bangalore, 560012.\protect\\
E-mail: \{raghavendrag,vijayn\}@iisc.ac.in}}

%
%

\markboth{~}%
{Sridharamurthy and Natarajan: Comparative Analysis of Merge Trees using Local Tree Edit Distance}
%



\IEEEtitleabstractindextext{%
\begin{abstract}
Comparative analysis of scalar fields is an important problem with various applications including feature-directed visualization and feature tracking in time-varying data. Comparing topological structures that are abstract and succinct representations of the scalar fields lead to faster and meaningful comparison. While there are many distance or similarity measures to compare topological structures in a global context, there are no known measures for comparing topological structures locally. While the global measures have many applications, they do not directly lend themselves to fine-grained analysis across multiple scales. We define a local variant of the tree edit distance and apply it towards local comparative analysis of merge trees with support for finer analysis. We also present experimental results on time-varying scalar fields, 3D cryo-electron microscopy data, and other synthetic data sets to show the utility of this approach in applications like symmetry detection and feature tracking.
\end{abstract}

}

\maketitle

\IEEEdisplaynontitleabstractindextext

%
\IEEEpeerreviewmaketitle

\input{section1.tex}
\input{section2.tex}
\input{section3.tex}
\input{section4.tex}
\input{section5.tex}
\input{section6.tex}
\input{section7.tex}


%

%
%
%
\ifCLASSOPTIONcompsoc
\section*{Acknowledgments}
\else
\section*{Acknowledgment}
\fi
We thank Tino Weinkauf for providing us with the 3D vortex street dataset with Okubo-Weiss criterion in full temporal resolution.
This work is supported by a Swarnajayanti Fellowship from the Department of Science and Technology, India (DST/SJF/ETA-02/2015-16), a scholarship from MHRD, Govt. of India, and a Mindtree Chair research grant.



\bibliographystyle{IEEEtran}
\typeout{}
\bibliography{localsimilarity}

\vskip -3\baselineskip plus -1fil

\begin{IEEEbiography}[{\includegraphics[width=1in,height=1.25in,clip,keepaspectratio]{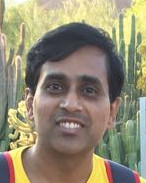}}]{Raghavendra Sridharamurthy} is a PhD candidate in computer science at Indian Institute of Science, Bangalore. He received BE degree in information technology from National Institute of Technology Karnataka, Surathkal and M.Sc degree in computer science from Indian Institute of Science. His research interests include scientific visualization, computational topology and its applications.
\end{IEEEbiography}

\vskip -3\baselineskip plus -1fil

\begin{IEEEbiography}[{\includegraphics[width=1in,height=1.25in,clip,keepaspectratio]{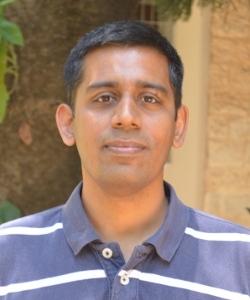}}]{Vijay Natarajan} is the Mindtree Chair Professor in the Department of Computer Science and Automation at Indian Institute of Science, Bangalore. He received the Ph.D. degree in computer science from Duke University. His research interests include scientific visualization, computational topology, and computational geometry. In current work, he is developing topological methods for time-varying and multi-field data visualization, and studying applications in biology, material science, and climate science.
\end{IEEEbiography}

\end{document}

%% file: section1.tex
\section{Introduction}

\begin{figure}
\subfigure[2D scalar field]{\label{fig:domain}\includegraphics[height=1.14in]{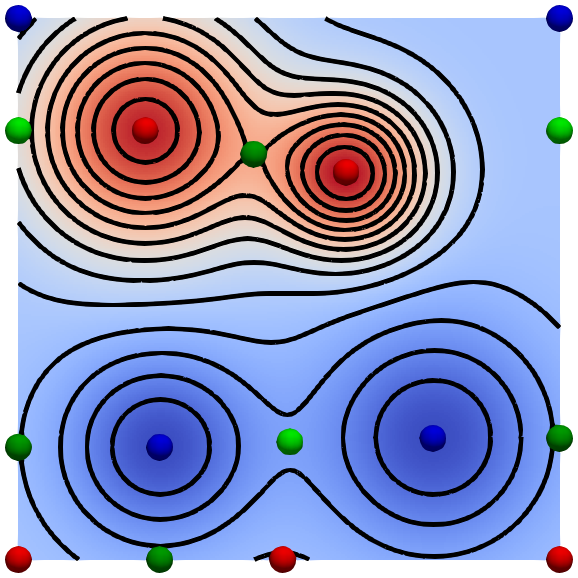}}
\subfigure[join tree]{\label{fig:join}\includegraphics[height=1.14in]{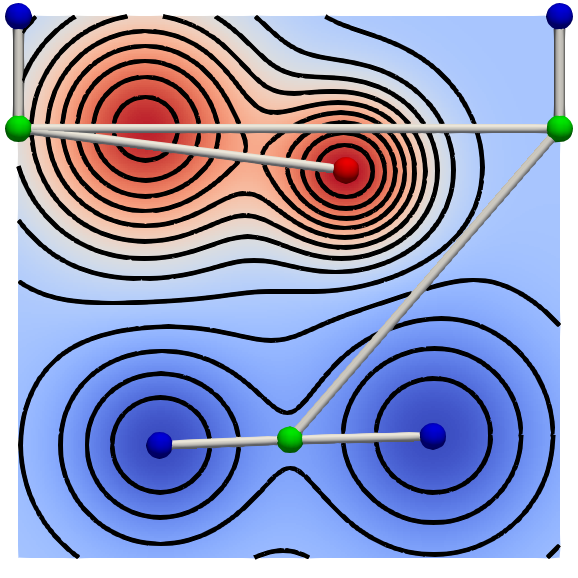}}
\subfigure[split tree]{\label{fig:split}\includegraphics[height=1.14in]{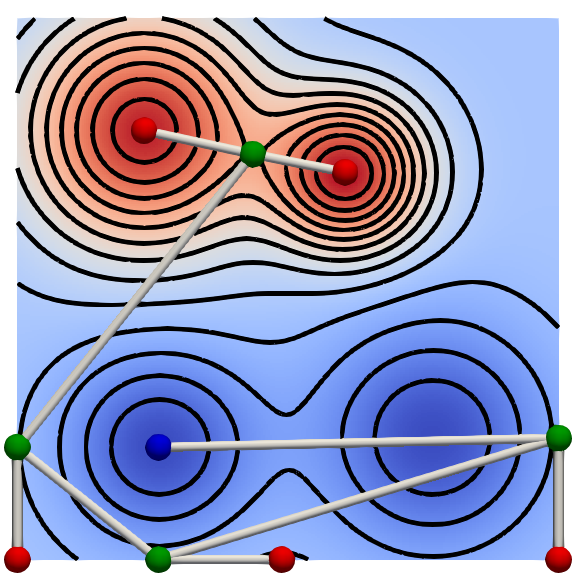}}
\caption{Merge trees. (a)~A 2D scalar field together with its critical points and a set of isocontours. (b,c)~A merge tree tracks the connectivity of sublevel sets (preimage $f^{-1} (-\infty, c]$) or the superlevel sets (preimage $f^{-1} [c,\infty)$). We consistently use the (\protect\includegraphics[height=0.15cm]{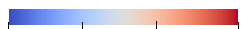}) color map for the scalar field and the (\protect\includegraphics[height=0.3cm]{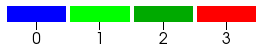}) color map for representing critical points based on their Morse index (0: minimum, 1: 1-saddle, 2: 2-saddle, 3: maximum).}
\label{fig:mergetree}
\vspace{-0.15in}
\end{figure}

\begin{figure}
\vspace{-0.05in}
\centering
\subfigure[Split tree 1]{\label{fig:sp1}\includegraphics[width = 0.2\textwidth]{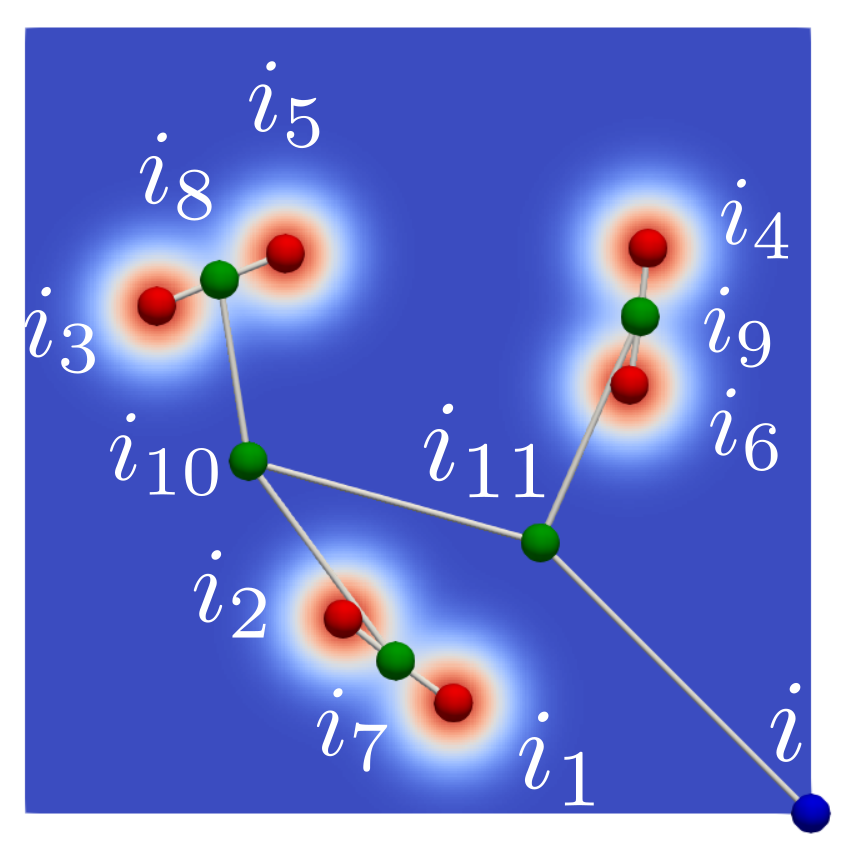}}
\subfigure[Split tree 2]{\label{fig:sp2}\includegraphics[width = 0.2\textwidth]{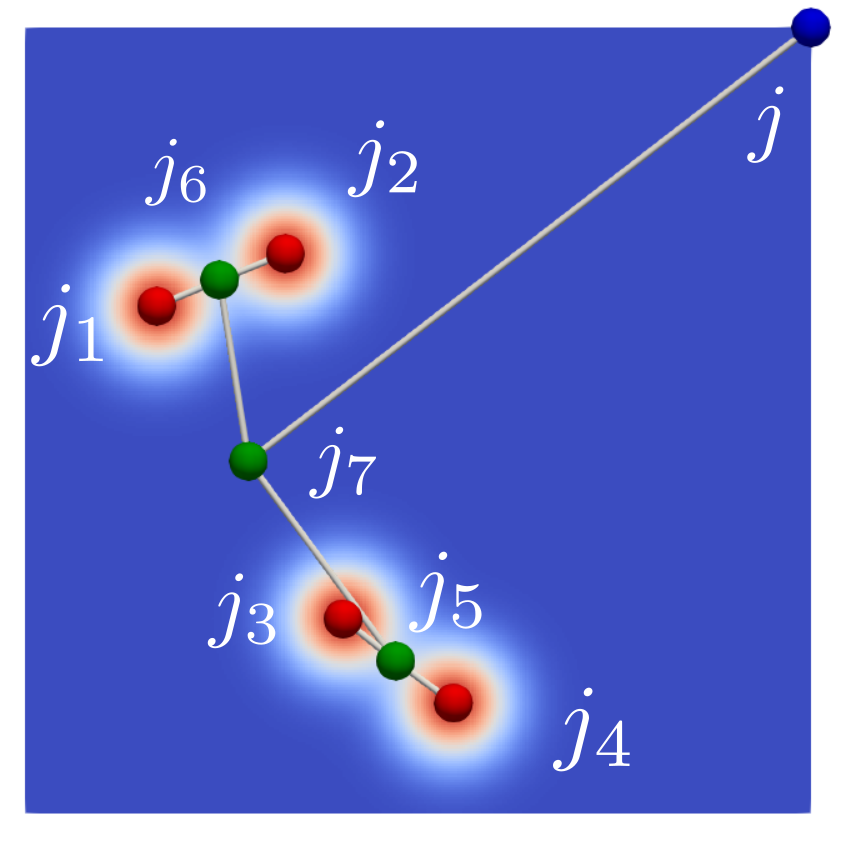}}
\caption{Scalar fields $f_1$ and $f_2$ that are not globally similar but contain locally similar regions. Split trees are overlaid on top of the scalar fields}
\label{fig:local}
\vspace{-0.15in}
\end{figure}

Comparative analysis and visualization of scalar fields is an  important problem with applications to feature detection and tracking, symmetry detection in scalar fields, and in general to the study of time-varying data. Topological structures like merge trees (Figure~\ref{fig:mergetree}) provide an abstract and combinatorial representation of the scalar field. These representations enable the analysis methods to focus on topological features of interest. A careful study of similarities and differences between the topology-based representations can lead to meaningful comparisons of the underlying scalar fields. Multiple similarity measures (alternatively comparison measures or distance measures) have been proposed to compare scalar fields and topological structures~\cite{cohen2007,di2012,morozov2013,bauer2014,beketayev2014,Saikia2014,saikia2015,dey2015,narayanan2015,di2016,saikia2017,Sridh2017,Sridh2020}. However, most of them describe methods to compare these structures globally. While global comparisons help us address a variety of interesting problems such as feature tracking, detection of periodicity in time-varying data, shape comparison, and temporal summarization, it cannot be used for finer analysis, specifically of the local structure or substructures of the corresponding topological features. For example, consider the split trees rooted at $i$ and $j$ in Figures~\ref{fig:sp1} and~\ref{fig:sp2}. Global comparison measures convey the overall dissimilarity, but do not capture the similarity between the regions that map to the pairs of subtrees rooted at $(i_{10},j_7)$ , $(i_8,j_6)$ and $(i_7,j_5)$ respectively.  This type of fine grained or multi-scale analysis leads to interesting applications and hence there is a need for a measure that detects similarity both locally and across multiple scales.

\subsection{Contributions}
In this paper, we propose a local tree edit distance based method to compare substructures of scalar fields across multiple scales. The comparison measure is an adaptation of the global tree edit distance for merge trees (\MTED) introduced by Sridharamurthy et al.~\cite{Sridh2020}. However, it is substantially different in terms of the definition, properties, approach to its computation, and applications. This paper makes the following key contributions:  
\begin{enumerate}
\item A novel local tree edit distance (\LMTED) to compare substructures in scalar fields.
\item A proof that it satisfies metric properties.
\item A dynamic programming algorithm to compute the \LMTED efficiently.
\item A notion of truncated persistence to compute costs of matching / correspondences, which brings in the additional benefit of saving computation time by reducing the number of comparisons.
\item Experiments to demonstrate the practical value of the distance towards symmetry detection at multiple scales, analysis of the effects of smoothing and subsampling, a fine grained analysis of topological compression, and applications to feature tracking. 
\end{enumerate}
The \MTED supports only a few of the above-mentioned applications. Even in these cases, it is restricted to comparisons on a coarser level or requires a higher level of user intervention. Feature tracking is not possible with \MTED without significant modifications.
\vspace{-0.20in}
\subsection{Related work}
Point-to-point comparisons like \RMS distance or norms such as $L^p, 1 \le p < \infty$ may be used to compare scalar fields. But, there is no natural notion of hierarchy that can be harnessed. Further, these measures are sensitive to small perturbations in the data. 

Existing measures that compare topological structures, and by extension the underlying scalar fields, are global. They sometimes ignore the hierarchy present in topological structures such as the contour tree and typically cannot be used to compare local structures in a meaningful way. Sridharamurthy et al.~\cite{Sridh2020} summarize previously developed comparison measures and describe in detail several methods including the bottleneck distance between persistence diagrams~\cite{cohen2007}, Interleaving distance~\cite{morozov2013}, distance based on branch decomposition of merge trees~\cite{beketayev2014}, edit distance between Reeb graphs~\cite{di2016,di2012}, functional distortion distance between Reeb graphs~\cite{bauer2014}, and comparing metric graphs by persistence distortion~\cite{dey2015}. Many of these methods are computationally intractable or they are setup in such a way that the computation considers the topological structure as a whole without harnessing the hierarchy present within.

Simple and practical similarity measures that are not metrics have also been studied.  Saikia et al.~\cite{saikia2015} propose a measure that compares histograms constructed based on the merge trees. As in the case of bottleneck distance, this measure ignores the topological structure during comparison. The applications do involve local structures but they are query based. The measure can be computed efficiently and is useful in practice. Saikia and Weinkauf~\cite{saikia2017} extend this measure and demonstrate applications to feature tracking in time-varying data. The measure is used to track a single feature (or a collection of features) if the feature is specified by the user in a query. 

Saikia et al.~\cite{Saikia2014} also introduce the extended branch decomposition graph (eBDG) that describes a hierarchical representation of all subtrees of a merge tree and designed an efficient algorithm to compare them. They also present experimental results on time-varying data. While the representation and comparison is based on the hierarchy, they demonstrate its use in applications by explicitly choosing or selecting region(s) of interest rather than considering the collection of all pairs of subtrees.

Scalar fields may also be compared based on their isosurfaces. Since theoretically, the number of isosurfaces is infinite, the comparison requires a selection of a finite number of isosurfaces of interest. Tao et al.~\cite{tao2018} extend the notion of isosurface similarity maps, first conceptualized by Bruckner and M{\"o}ller~\cite{Bruckner2010}, to construct matrices of isosurface similarity maps (MISM), and use it to explore multivariate time-varying data. This involves construction of self-similarity maps, temporal similarity maps, and variable similarity maps followed by temporal clustering and variable grouping. Finally, paths spanning across these maps are used to guide the visual comparison. The choice of the isovalues used is crucial but the inclusion hierarchy followed by isosurfaces is not utilised.

Lukasczyk et al.~\cite{lukasczyk2017,lukasczyk2019} introduce the concept of nested tracking graphs to track the entire family of isosurfaces over time while preserving their nested hierarchy of the isosurfaces. Features are often represented by isosurfaces, so the method applies to feature tracking. While the method does facilitate tracking features in all scales and across time, it does not support generic comparison between features. 

Symmetry detection in scalar fields is another important application that involves the comparison of local substructures to decide if the scalar field contains repeating patterns. The notion of symmetry has been well studied by Thomas and Natarajan~\cite{thomas2011, thomas2013,thomas2014}. While the comparison methods that they describe work well for symmetry detection, the methods have not been applied to detect local similarities between different scalar fields in general. In other words, the symmetry detection problem is a special case of the more general local similarity detection problem.

 Sridharamurthy et al.~\cite{Sridh2020} introduce a tree edit distance between merge trees (\MTED). The algorithm to compute the distance processes the trees in a bottom-up fashion. Computing the global distance involves computing distances between various pairs of subtrees, which are not necessarily merge trees. In this work, we show how such a global tree edit distance can be extended towards a fine-grained comparison of scalar fields. Specifically, the global \MTED is restricted to cases where only the tree at the top of the hierarchy is guaranteed to be a merge tree. In contrast, we facilitate comparison between all pairs of subtrees of merge trees by ensuring that all comparisons are between trees that are guaranteed to be merge trees.

%% file: section2.tex
\section{Tree Edit Operations}
In this section, we introduce necessary background on edit operations between rooted trees with a focus on merge trees.

\subsection{Merge tree}
\label{sec:mergetree}
Let $f: \mathbb{X} \longrightarrow \mathbb{R}$ denote a scalar function defined on a manifold domain $\mathbb{X}$. A \emph{level set} is the preimage $f^{-1} (c)$ of a real value $c$ (Figure~\ref{fig:domain}). The merge tree of $f$ (Figures~\ref{fig:join} and~\ref{fig:split}) tracks the connectivity of sublevel sets ($f^{-1} (-\infty, c]$, \emph{join tree}) or superlevel sets ($f^{-1} [c,\infty)$, \emph{split tree})~\cite{carr2003}. 
The split tree (Figure~\ref{fig:spt}) of a generic scalar function is a simple rooted binary tree. Nodes of the split tree include critical points of $f$: maxima, saddles, and the global minimum (root). The maxima have 0 children, the saddles have 2 children, and the global minimum has 1 child. All maxima are paired with saddles based on the notion of topological persistence~\cite{edelsbrunner2000}, except for the global maximum which is paired with the global minimum. A persistence pair $(m,s)$ represents a topological feature (connected component of superlevel set) that is created at a maximum $m$ during a downward sweep of the domain $\mathbb{X}$ and destroyed at a saddle $s$. The \emph{persistence} of such a maximum-saddle pair is defined as the difference in function value at the two critical points, $pers(m) = pers(s) = f(m) - f(s)$. The \emph{persistence diagram} is a plot of the persistence pairs $(f(m),f(s))$ on the plane. The join tree is defined similarly, its nodes consist of minima, saddles, and the global maximum (root). Several serial and parallel algorithms are available for fast computation of merge trees~\cite{carr2003,acharya2015parallel,Morozov2013dist}, \cite{Gueunet2017}.

\subsection{Tree edit distance}
The distance between a pair of trees may be defined by introducing edit operations that transform one tree into another~\cite{Bille2005}. Each edit operation has an associated cost. The tree edit distance is defined as the cost associated with a sequence of edit operations that transforms one tree into another while minimizing the total cost. We now introduce the edit operations between labeled trees, following definitions from Zhang~\cite{Zhang1996}. For a node $i$ in a rooted tree $T$, $deg(i)$ denotes the number of children of $i$ and $parent(i)$ is its parent in the tree. The maximum degree of a node in $T$ is denoted as $deg(T)$. The set of labels of $T$ is denoted by $\Sigma$, and $\lambda \notin \Sigma$ denotes the null or empty character. An empty tree is denoted by $\theta$. Since we focus on labeled trees, we use the same notation to refer to the node and its label. 

\myparagraph{Edit operations.}
\label{sec:editcost}  
We consider three edit operations that modify a rooted tree, one node at a time: \emph{relabel, insert,} and \emph{delete}, together with associated costs.

\begin{enumerate}
\item \textbf{relabel:} A relabel $i \longrightarrow j$ corresponds to an operation where the label $i \in \Sigma$ of a node is changed to a label $j \in \Sigma$.
\item \textbf{delete:} A delete operation $i \longrightarrow \lambda$ removes a node with label $i \in \Sigma$ and all children of the node $i$ are made children of $parent(i)$. 
\item \textbf{insert:} An insert operation $\lambda \longrightarrow j$ inserts a node with label $j \in \Sigma$ as a child of another node $i$ by moving all children of $i$ to children of $j$.
\end{enumerate}

We define a cost function $\gamma$ that assigns a non-negative real number to each edit operation. It is useful if the cost function $\gamma$ satisfies metric properties~\cite{Sridh2020}.

\myparagraph{Edit distance.} 
The distance between two trees $T_1, T_2$ is defined as
\begin{align}
D_e(T_1,T_2) = \min_{S}\{\gamma(S)\},
\end{align}
where $S$ is a sequence of edit operations that transforms $T_1$ to $T_2$. Zhang et al.~\cite{Zhang1992} showed that $D_e$ is equal to the cost of an ancestor preserving mapping between nodes of $T_1$ and $T_2$ but computing $D_e$ is an NP-complete problem.

\subsection{Constrained edit distance mappings}
\label{sec:constrainedmapping}
Incorporating an additional constraint, namely restricting the mapping of disjoint subtrees to disjoint subtrees, makes the problem computationally tractable. This constraint is clearly meaningful in the context of merge trees because disjoint subtrees correspond to spatially disjoint regions in the domain and hence spatially disjoint features. 
We now describe a constrained edit distance $D_c$ and the corresponding mappings~\cite{Zhang1996}. The recursive definition of $D_c$ naturally incorporates the constraint. We refer the reader to the supplementary material for a few illustrative examples. 

We use $i$ and $j$ with or without subscripts to denote both the nodes and their labels in the trees $T_1$ and $T_2$, respectively. Let $i_1, i_2, \ldots, i_{n_i}$ be the children of $i$ and $j_1, j_2, \ldots, j_{n_j}$ be the children of $j$.  Let $T[i]$ denote the subtree rooted at $i$ and $F[i]$ denote the unordered forest obtained by deleting the node $i$ from $T[i]$. Further, let $\theta$ denote the empty tree. Then, 
\begin{align}
D_c(\theta,\theta) &= 0, \\
D_c(F_1[i],\theta) &= \sum\limits_{k=1}^{n_i} D_c(T_1[i_k],\theta), \\
D_c(T_1[i],\theta) &= D_c(F_1[i],\theta) + \gamma(i \longrightarrow \lambda),\\
D_c(\theta,F_2[j]) &= \sum\limits_{k=1}^{n_j} D_c(\theta,T_2[j_k]), \\
D_c(\theta,T_2[j]) &= D_c(\theta,F_2[j]) + \gamma(\lambda \longrightarrow j),
\end{align}
\begin{align}
&\scriptsize D_c(T_1[i],T_2[j]) \nonumber \\
&= \scriptsize \min \begin{cases}
D_c(\theta,T_2[j]) + \min\limits_{1 \le t \le n_j} \{D_c(T_1[i], T_2[j_t])- D_c(\theta, T_2[j_t])\},\\
D_c(T_1[i],\theta) + \min\limits_{1 \le s \le n_i} \{D_c(T_1[i_s], T_2[j])- D_c(T_1[i_s],\theta)\},\\
D_c(F_1[i], F_2[j]) + \gamma(i \longrightarrow j).
\end{cases}
\end{align}
\begin{align}
&\scriptsize D_c(F_1[i],F_2[j]) \nonumber \\
&= \scriptsize \min \begin{cases}
D_c(\theta,F_2[j]) + \min\limits_{1 \le t \le n_j} \{D_c(F_1[i], F_2[j_t])- D_c(\theta, F_2[j_t])\},\\
D_c(F_1[i],\theta) + \min\limits_{1 \le s \le n_i} \{D_c(F_1[i_s], F_2[j]) - D_c(F_1[i_s],\theta)\},\\
\min\limits_{M_r(i,j)} \gamma(M_r(i,j)).
\end{cases}
\end{align}
Here, $M_r(i,j)$ is the restricted edit distance mapping between $F_1[i]$ and $F_2[j]$. Nodes within different trees of $F_1$ are mapped to nodes lying in different trees of $F_2$. Essentially, if $M_r$ maps node $i_1$ to $j_1$ and node $i_2$ to $j_2$, then $i_1$ and $i_2$ belong to a common tree in $F_1[i]$ if and only if $j_1$ and $j_2$ belong to a common tree in $F_2[j]$. The minimum cost restricted mapping may be computed by constructing a weighted bipartite graph (Figure~\ref{fig:moddist}) in such a way that the cost of the minimum weight maximum matching $MM(i,j)$ is exactly the same as the cost of the minimum restricted mapping $M_r(i,j)$,
\begin{align}
\min\limits_{M_r(i,j)} \gamma(M_r(i,j)) = \min\limits_{MM(i,j)} \gamma(MM(i,j))
\end{align}

\subsection{\MTED between merge trees}
\label{sec:teddefinition}

The constrained tree edit distance (\MTED) between two merge trees $T_1, T_2$ is defined as the constrained edit distance
\begin{align}
\MTED (T_1, T_2) = D_c(T_1,T_2)
\end{align}
where the cost of edit operations is governed by one of the two cost models (the $L^\infty$ cost $C_W$ or overhang cost $C_O$) introduced by Sridharamurthy et~al.~\cite{Sridh2020} instead of the generic cost model defined by Zhang~\cite{Zhang1996}. 

Both the $L^\infty$ cost $C_W$ and the overhang cost $C_O$, are intuitive and incorporate properties of merge trees. For the sake of illustration, we describe $C_W$ here and refer to \cite[Section 4.2.2]{Sridh2020} for the definition of $C_O$. 

Consider nodes $p \in T_1$ and $q \in T_2$. Both $p$ and $q$ appear as points $(b_p, d_p)$ and $(b_q, d_q)$ in the respective persistence diagrams. Here, $b_p$ represents the lower of the function values between $p$ and its persistence pair, also referred to as the birth time of the topological feature. Similarly, $d_p$ represents the higher value, also referred as the death time. The $C_W$ cost of edit operations are defined as
\begin{align}
\gamma(p \longrightarrow q) &= \min \begin{cases}
\max(|b_q-b_p|,|d_q-d_p|),\\
\frac{(|d_p-b_p| + |d_q-b_q|)}{2}
\end{cases}\\
\gamma(p \longrightarrow \lambda) &= \frac{|d_p-b_p|}{2}\\
\gamma(\lambda \longrightarrow q) &= \frac{|d_q-b_q|}{2}
\end{align}
The constrained edit distance $D_c$ with the $C_W$ cost model can be used to compute \MTED using a dynamic programming algorithm. 

%% file: section3.tex
\section{Local Tree Edit Distance}
\label{sec:lted}
\begin{figure}
\vspace{-0.05in}
\subfigure[Split tree 1]{\label{fig:spt1}\includegraphics[width = 0.28\textwidth]{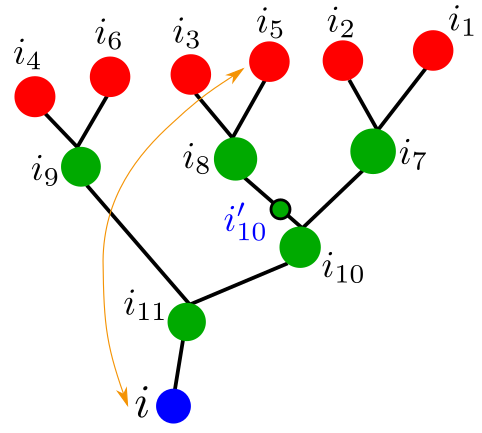}}
\subfigure[Split tree 2]{\label{fig:spt2}\includegraphics[width = 0.20\textwidth]{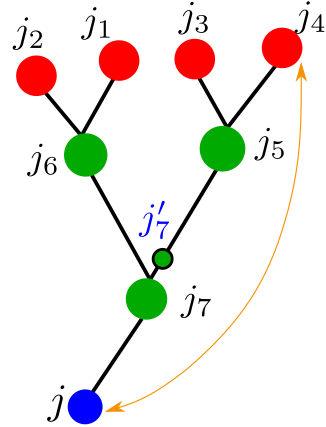}}
\caption{Split trees  corresponding to the scalar fields shown in Figure~\ref{fig:local}. One persistence pair is shown within each split tree (orange link). When comparing subtrees $T_1[i_8]$ or $T_2[j_5]$, nodes $i_5$ and $j_4$ are unpaired. So, dummy nodes $i'_{10}$ and $j'_7$ are inserted with $|f_1(i_{10}) - f_1(i'_{10})| < \varepsilon$ and $|f_2(j_7) - f_2(j'_7)| < \varepsilon$ to serve as root and as a pair of the unpaired node.}
\label{fig:spt}
\vspace{-0.15in}
\end{figure}

We introduce a local version of $D_c$, a local tree edit distance (\LMTED) between merge trees. We begin with a few necessary definitions. For illustration, we consider the split trees in Figure~\ref{fig:spt}. These trees correspond to the scalar fields in Figure~\ref{fig:local}.
\subsection{Truncated persistence}
\label{sec:truncpers}
The cost of edit operations in \MTED~\cite{Sridh2020} is based on the topological persistence of the node(s). Specifically, their $L^{\infty}$ cost $C_W$ is the $L^{\infty}$ distance between the point pair in the persistence diagram corresponding to the two nodes (relabel) or the distance between the point in the persistence diagram and the diagonal (insert / delete). While comparing subtrees such as $T_2[j_5]$ (Figure~\ref{fig:spt2}), a node $j_4$ whose persistence pair is the global root $j$ contributes a large value to the edit distance. Using the same cost for $j_4$ while comparing all subtrees containing the node may not be appropriate. For example, subtrees $T_2[j_5]$ and $T_2[j_{}]$ map to two regions in the domain that are vastly different in size. To alleviate this inconsistency,  we introduce the notion of dummy nodes and truncated persistence. While comparing the subtree $T_2[j_5]$ with other subtrees, we insert a dummy node $j'_7$ that serves as a root of the subtree $T_2[j_5]$ and as the pair of $j_4$. The function value at the dummy node differs from the parent of $j_5$ at most by a small value $\varepsilon$.

An unpaired node in a subtree corresponds to a leaf whose persistence pair is outside the subtree. We use $i_u$ and $j_u$ to denote unpaired nodes in $T_1[i]$ and $T_2[j]$, respectively. Dummy nodes corresponding to $T_1[i]$ and $T_2[j]$ are denoted by $i'$ and $j'$. For an unpaired node $i_u \in T_1[i]$ and a truncated root that is represented by a dummy node $i'$, we define \emph{truncated persistence} as
\begin{align}
tp_{i'}(i_u) = |f_1(i_u)-f_1(i')|.
\end{align}
Note that a leaf node will be unpaired in some subtree and, in some cases, it can be unpaired in multiple subtrees (for example, $j_4$). The subscript $i'$ in the definition specifies the subtree under consideration.

\subsection{Truncated cost} 
The cost of the edit operations need to be updated based on the  truncated persistence for unpaired nodes $i_u$ and $j_u$. Let $\gamma$ denote the original cost of an edit operation derived from the $L^{\infty}$ cost $C_W$ used in \MTED. We define a new truncated cost $\gamma'$ as
\begin{align}
    \gamma'(i \longrightarrow j) = \begin{cases} 
    \gamma(i \longrightarrow \lambda), i \ne i_u, j = j_u|\lambda, \\
    \gamma(\lambda \longrightarrow j), i = i_u|\lambda, j \ne j_u, \\
    \gamma(i \longrightarrow j), i \ne i_u, j \ne j_u, \\
    0, i = i_u|\lambda, j = j_u|\lambda,
    \end{cases}
    \label{eqn:truncatededitcost}
\end{align}

\subsection{\LMTED between merge trees}
We now describe a new local tree edit distance that is appropriate for localized comparison of merge trees, discuss its properties, and present an algorithm for computing the distance. The local tree edit distance (\LMTED) for a pair of trees rooted at $i$ and $j$ is denoted $\LMTED(i,j)$ and defined as follows:
\begin{align}
\LMTED(i,j) = D'(i,j) + \Gamma(i_u \longrightarrow j_u).
\end{align}
Here, $\Gamma(i_u \longrightarrow j_u)$ denotes the relabel cost computed using the truncated persistence values of $i_u$ and $j_u$. $D'(i,j)$ is the modified edit distance between the two trees that excludes the cost between the unpaired nodes from each tree. 

\begin{figure}
    \centering
    \includegraphics[width=0.45\textwidth]{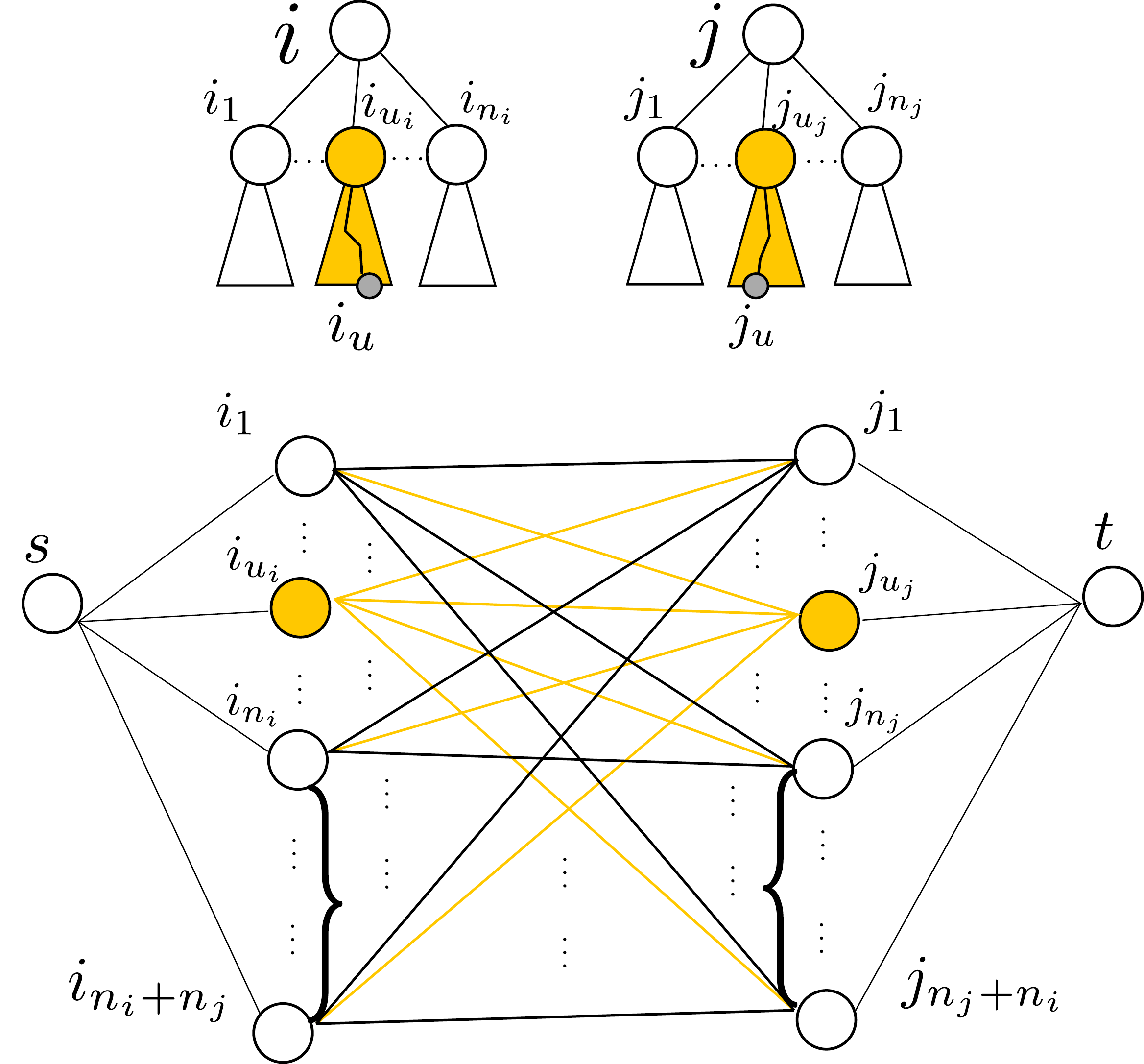}
    \caption{Illustrating \LMTED. To compute \LMTED between subtrees rooted at $i$ and $j$, we treat the subtrees containing the unpaired nodes $i_u,j_u$, labeled as $i_{u_i},j_{u_j}$  and highlighted in orange, differently. For other nodes, the formulation is same as \MTED. For $i_u,j_u$, we use truncated persistence to determine the costs. In the matching required to compute $M'_r(i,j)$, we consider truncated persistence to determine the weights of all edges incident on $i_{u_i},j_{u_j}$ (highlighted in orange). $s$ and $t$ are the source and destination nodes of the flow problem that is equivalent to the matching problem to determine $M'_r(i,j)$. The cardinalities of the two sides are made equal by inserting a set of $n_j$ dummy nodes adjacent to $s$ and $n_i$ dummy nodes adjacent to $t$.
    }
    \label{fig:moddist}
    \vspace{-0.15in}
\end{figure}

We now describe the recursive formulation of $D'(i,j)$. The key difference between \LMTED and \MTED is the way in which unpaired nodes are handled. The original formulation (as for $D_c$) applies as is for the paired nodes, not for the unpaired nodes. The unpaired node changes depending on the level of the subtree, as we move from the leaves towards the root of the merge tree. As we traverse a merge tree bottom-up considering all possible subtrees, a node can be unpaired until we reach the level containing its pair, following which it remains paired until the end. When the node is unpaired, at every level its contribution changes. We account for the contribution from the unpaired node to $D'(i,j)$ in a final step and so we are able to retain a recursive formulation. Once the node is paired, its contribution (equal to its persistence) does not change further. This demands a new recursive formulation which can handle both the scenarios. The new formulation $D'$ will thus be a modification of $D_c$.

From Figure~\ref{fig:moddist} and using similar terminology as before, let $i_1, i_2, \ldots, i_{n_i}$ be the children of $i$ and $j_1, j_2, \ldots, j_{n_j}$ be the children of $j$.  Let $T_1[i]$ denote the subtree rooted at $i$ and $F_1[i]$ denote the unordered forest obtained by deleting the node $i$ from $T_1[i]$. Again, $i_u$ and $j_u$ are unpaired nodes in $T_1[i]$ and $T_2[j]$, respectively. Let $i_{u_i}$ be the child lying on the path between $i$ and $i_u$ in $T_1[i]$ and $j_{u_j}$ be the child lying on the path between $j$ and $j_u$ in $T_2[j]$.

Recall that $\theta$ denotes the empty tree and $D_c(,)$ denotes \MTED. Then $D'$ is recursively defined as follows:
\vspace{-0.15in}
\begin{align}
D'(\theta,\theta) &= 0, \\
D'(F_1[i],\theta) &= \sum\limits_{k=1, k \neq u_i}^{n_i} D_c(T_1[i_k],\theta) + D'(T_1[i_{u_i}],\theta), \\
D'(T_1[i],\theta) &= D'(F_1[i],\theta) + \gamma'(i \longrightarrow \lambda),\\
D'(\theta,F_2[j]) &= \sum\limits_{k=1, k \neq u_j}^{n_j} D_c(\theta,T_2[j_k]) + D'(\theta,T_2[j_{u_j}]), \\
D'(\theta,T_2[j]) &= D'(\theta,F_2[j]) + \gamma'(\lambda \longrightarrow j),
\end{align}

\begin{align*}
min_{T_2} = \min \begin{cases}\min\limits_{1 \le t \le n_j, t \neq u_j} \{D_c(T_1[i], T_2[j_t])- D_c(\theta, T_2[j_t])\},\\
\{D'(T_1[i], T_2[j_{u_j}])- D'(\theta, T_2[j_{u_j}])\}
\end{cases}
\end{align*}

\begin{align*}
min_{T_1} = \min \begin{cases}\min\limits_{1 \le s \le n_i, t \neq u_i} \{D_c(T_1[i_s], T_2[j])- D_c(T_1[i_s],\theta)\},\\
\{D'(T_1[i_{u_i}], T_2[j])- D'(T_1[i_{u_i}],\theta)\}
\end{cases}
\end{align*}

\begin{align*}
min_{F_j} = \min \begin{cases}\min\limits_{1 \le t \le n_j, t \neq u_j} \{D_c(F_1[i], F_2[j_t])- D_c(\theta, F_2[j_t])\},\\
\{D'(F_1[i], F_2[j_{u_j}])- D'(\theta, F_2[j_{u_j}])\}
\end{cases}
\end{align*}

\begin{align*}
min_{F_i} = \min \begin{cases}\min\limits_{1 \le s \le n_i, t \neq u_i} \{D_c(F_1[i_s], F_2[j])- D_c(F_1[i_s],\theta)\},\\
\{D'(F_1[i_{u_i}], F_2[j])- D'(F_1[i_{u_i}],\theta)\}
\end{cases}
\end{align*}
\begin{align}
D'(T_1[i],T_2[j]) &= \min \begin{cases}
D'(\theta,T_2[j]) + min_{T_2},\\
D'(T_1[i],\theta) + min_{T_1},\\
D'(F_1[i], F_2[j]) + \gamma'(i \longrightarrow j).
\end{cases}
\end{align}

\begin{align}
D'(F_1[i],F_2[j]) &= \min \begin{cases}
D'(\theta,F_2[j]) + min_{F_j},\\
D'(F_1[i],\theta) + min_{F_i},\\
\min\limits_{M'_r(i,j)} \gamma'(M'_r(i,j)).
\end{cases}
\end{align}

All terms that involve subtrees containing the unpaired node (orange) are updated to incorporate $D'$, whereas the constrained  edit distance $D_c$ appears elsewhere. The bipartite graph formulation that is used to compute the minimum cost restricted mapping between forests is also updated to incorporate $D'$ and is now denoted as $M'_r(i,j)$ (Figure~\ref{fig:moddist}, bottom).

\subsection{Cost model and properties}
We can employ either of the costs, the $L_\infty$ cost $C_W$ or the overhang cost $C_O$ introduced for the \MTED~\cite[Section $4.2$]{Sridh2020}. Both costs are proven to be metrics. If they remain so even with the newly introduced cost based on the truncated persistence, then by Zhang~\cite{Zhang1996} the \LMTED satisfies metric properties. The proofs of non-negativity and symmetry is straightforward because $\gamma'$ together with $\Gamma$ is defined in terms of $\gamma$, which in turn satisfies both properties because it is a combination of sum, max, and min of absolute values.

We now prove the triangle inequality for $\gamma'$ together with $\Gamma$. Let $T_1[i]$, $T_2[j],$ and $T_3[k]$ be three subtrees with unpaired nodes $i_u,j_u,k_u$. We insert dummy nodes $i',j',k'$ and construct trees $T_1[i']$, $T_2[j'],$ and $T_3[k']$. The truncated persistence values $tp_{i'} (i_u), tp_{j'} (j_u), tp_{k'} (k_u)$ in subtrees $T_1[i]$, $T_2[j], T_3[k]$ are respectively equal to the regular persistence values within trees $T_1[i']$, $T_2[j'], T_3[k']$. For a given triple $i_1 \in T_1[i], j_1 \in T_2[j], k_1 \in T_3[k]$, we will show that triangle inequality holds by considering different cases.\\ 
\textbf{Case 1:} Nodes $i_1,j_1,k_1$ are all unpaired or are all paired. When all the nodes $i_1,j_1,k_1$ are paired, $\gamma' = \gamma$ and hence triangle inequality holds. If all are unpaired,  $\gamma'(i_1 \longrightarrow j_1) = \gamma' (j_1 \longrightarrow k_1) = \gamma' (i_1 \longrightarrow k_1) = 0$. Further, $\Gamma(i_1 \longrightarrow j_1)$, $\Gamma(j_1 \longrightarrow k_1)$ and $\Gamma(i_1 \longrightarrow k_1)$ are equal to relabel costs ($\gamma$) for the trees $T_1[i']$, $T_2[j']$ and $T_3[k']$ and hence triangle inequality holds~\cite[Section $4.3$]{Sridh2020}. \\
\textbf{Case 2:} Two nodes are unpaired. The case where $i_1$ and $k_1$ are unpaired while $j_1$ is paired is impossible because of the constraint that unpaired nodes are mapped to unpaired nodes and the operation is forced to be a relabel. \\ 
\textbf{Case 2.1:} $i_1$ and $j_1$ are unpaired. Then the LHS
\begin{align}
    \gamma'(i_1 \longrightarrow j_1) + \gamma'(j_1 \longrightarrow k_1) + \Gamma(i_1 \longrightarrow j_1)\\
    = 0 + \gamma(\lambda \longrightarrow k_1) + \Gamma(i_1 \longrightarrow j_1)\\
    = \gamma(\lambda \longrightarrow k_1) + \Gamma(i_1 \longrightarrow j_1),
\end{align}
and the RHS  
\begin{align}
    \gamma'(i_1 \longrightarrow k_1) 
    = \gamma(\lambda \longrightarrow k_1).
\end{align}
Since $\Gamma(i_1 \longrightarrow j_1) \ge 0$ always, we have LHS $\ge$ RHS.\\
\textbf{Case 2.2:} $j_1$ and $k_1$ are unpaired, then the LHS 
\begin{align}
    \gamma'(i_1 \longrightarrow j_1) + \gamma'(j_1 \longrightarrow k_1) + \Gamma(j_1 \longrightarrow k_1)\\
    = \gamma(i_i \longrightarrow \lambda) + 0 + \Gamma(j_1 \longrightarrow k_1)\\
    = \gamma(i_i \longrightarrow \lambda) + \Gamma(j_1 \longrightarrow k_1),
\end{align}
and the RHS 
\begin{align}
    \gamma'(i_1 \longrightarrow k_1) 
    = \gamma(i_1 \longrightarrow \lambda).
\end{align}
Since $\Gamma(j_1 \longrightarrow k_1) \ge 0$ always, we again have LHS $\ge$ RHS.\\
\textbf{Case 3:} A single node is unpaired. This is impossible because of the constraint that unpaired nodes are mapped only to unpaired nodes via a relabel edit operation.\\
In all cases, the truncated cost $\gamma'$ together with $\Gamma$ satisfies the triangle inequality. So, it follows from Zhang~\cite{Zhang1996} that \LMTED is indeed a metric.

%% file: section4.tex
\section{Computing \LMTED}
We propose a modified dynamic programming based algorithm for computing \LMTED between merge trees. The use of truncated persistence as cost of the edit operations implies that \LMTED can be computed using solutions to non-overlapping sub-problems for computing $D_c$. However, the dynamic programming formulation needs to be modified because $D'$ recursively depends both on $D'$ and $D_c$. 

\subsection{Dynamic Programming tables}
Consider the subtrees $T[i_{10}]$ and $T[i_8]$ in Figure~\ref{fig:spt}. The node $i_5$ is unpaired. Within the subtree $T[i_8]$, node $i_5$ has truncated persistence $tp_{i'_8}(i_5)$. But, its truncated persistence is equal $tp_{i'_{10}}(i_5)$ when considering the subtree $T[i_{10}]$. Dynamic programming works by storing the results of sub-problems within a table so that it can be reused. Entries corresponding to both (in general, more than two) values of truncated persistence are required for the computation.

\begin{figure}
\centering
\includegraphics[width=0.25\textwidth]{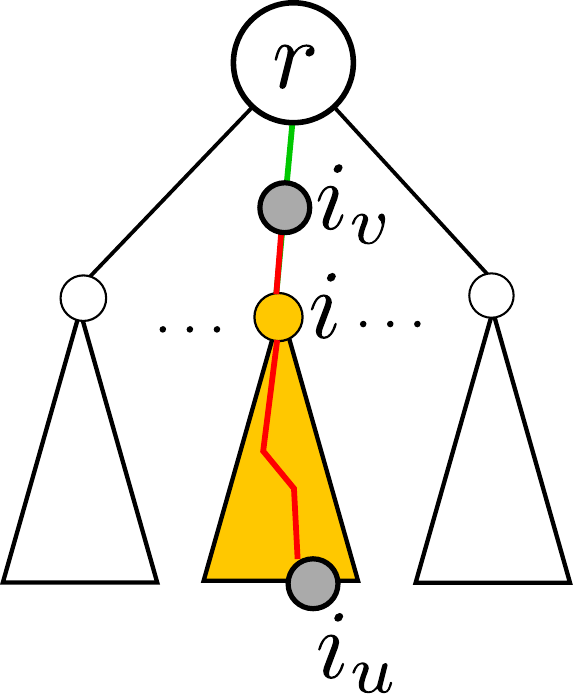}
\caption{Illustration of Cases O and M. The portion of the path colored in red corresponds to case M while the portion of the path colored in green corresponds to case O.}
\label{fig:illustration}
\vspace{-0.15in}
\end{figure}

Consider a subtree $T[i]$ with an unpaired node $i_u$ and the path from $i_u$ to the global root $r$ as shown in Figure~\ref{fig:illustration}. Let $i_{v}$ be the pair of $i_u$, clearly $i_{v} \in ancestor(i)$. While processing the unpaired node $i_u$, we need to distinguish between two cases:
\begin{enumerate}
    \item[\textbf{O}:] The contribution of $i_u$ is measured by persistence as defined in the usual sense.
    \item[\textbf{M}:] The contribution of $i_u$ is measured by an appropriate instance of truncated persistence. 
\end{enumerate} 
Case~\textbf{O} corresponds to all subtrees rooted at $i_a \in ancestor(i)$, where $i_{v} \le i_a \le r$ in the directed path highlighted in green in Figure~\ref{fig:illustration}. Case~\textbf{M} corresponds to trees rooted at $i_a$ such that $i_u \le i_a < i_{v}$, which also includes $i$ as highlighted in red in Figure~\ref{fig:illustration}.

\begin{figure}[t]
\centering
\includegraphics[width=0.32\textwidth]{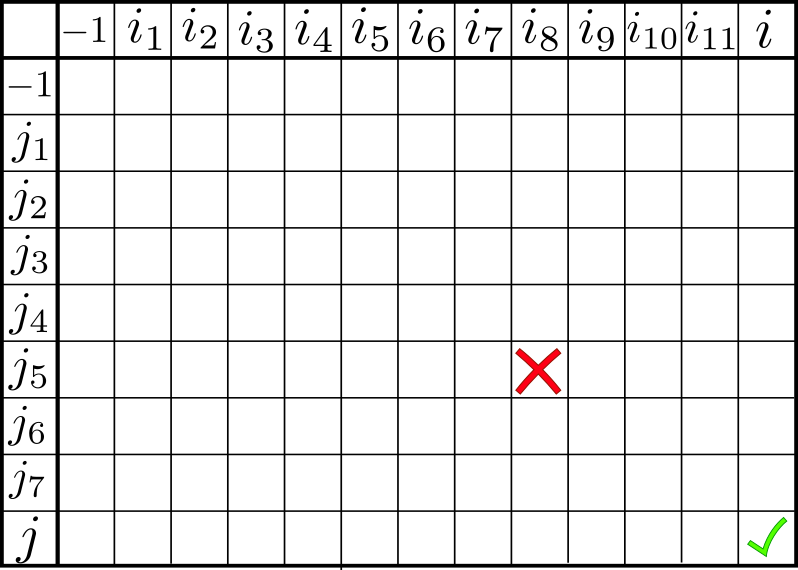}
\caption{Dynamic programming table for $D_c$. The only useful entry in this table is the one corresponding to $(i,j)$. Other entries do not represent a meaningful distance because the subtrees compared to compute the entries are not merge trees. Entries in the row and column labeled $-1$ correspond to comparison with the empty tree $\theta$.} 
\label{fig:origdptable}
\vspace{-0.05in}
\end{figure}

\begin{figure}
\centering
\includegraphics[width=0.45\textwidth]{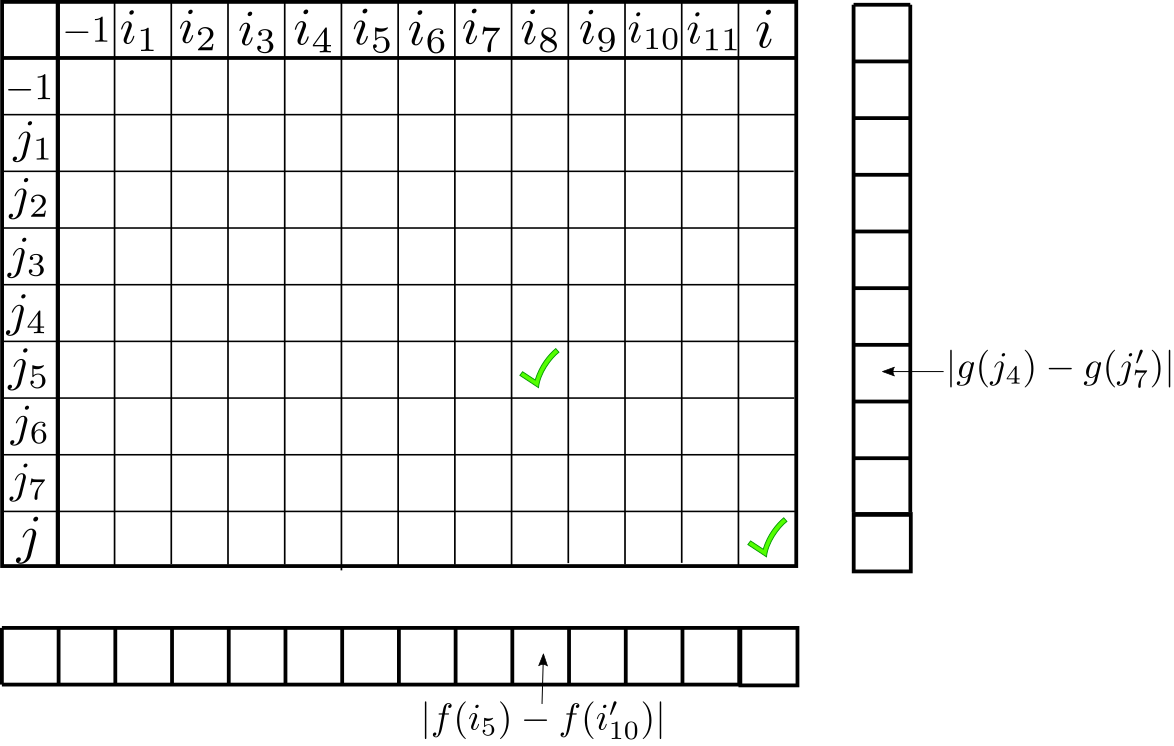}
\caption{The modified DP table corresponding to $D'$, which is used to compute \LMTED. This table contains an additional row and column when compared to the table for $D_c$. This additional row and column stores the truncated persistence values for the unpaired nodes corresponding to the current level. Entries in this table represent valid (and useful) distance between merge trees, which are subtrees of the trees rooted at $i$ and $j$, respectively.}
\label{fig:modifieddptable}
\vspace{-0.15in}
\end{figure}

\begin{figure}[t]
\centering
\includegraphics[width=0.32\textwidth]{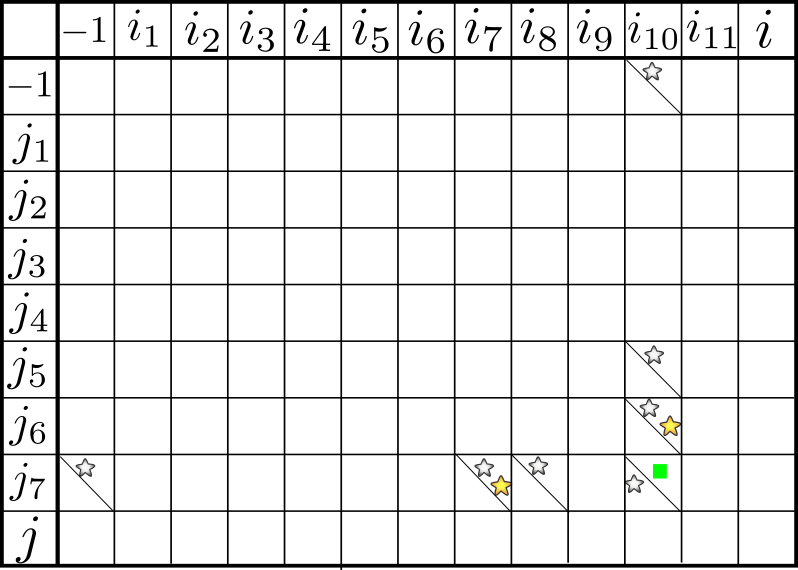}
\caption{Entries required to calculate $D_c(i,j)$ (\protect\includegraphics[height=0.15cm]{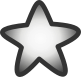}) and $D'(i,j)$ (\protect\includegraphics[height=0.15cm]{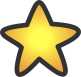}) in the table for $D_c$. The required entry is denoted by a green square.The lower triangle entries correspond to forests and upper triangle entries to trees.}
\label{fig:origent}
\vspace{-0.1in}
\end{figure}

\begin{figure}
\centering
\includegraphics[width=0.35\textwidth]{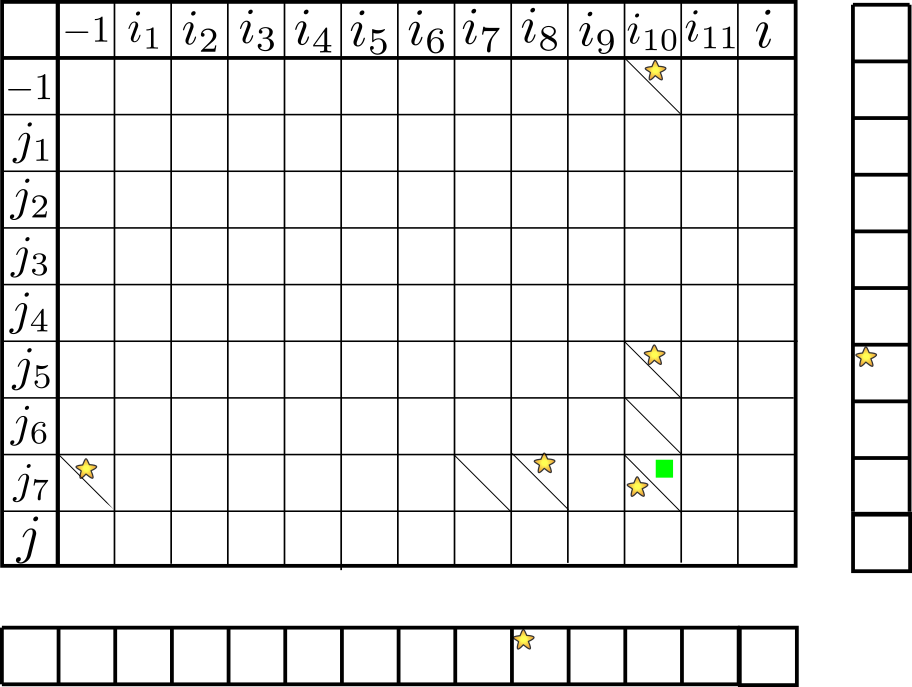}
\caption{Entries required to calculate $D'(i,j)$ (\protect\includegraphics[height=0.15cm]{Images/star2.png}) in the modified table. The required entry is denoted by a green square.The lower triangle entries correspond to forests and upper triangle entries to trees.}
\label{fig:modent}
\vspace{-0.15in}
\end{figure}

We wish to design an algorithm that makes a single pass over the two input merge trees and computes \LMTED between all pairs of subtrees. In order to achieve this objective, we propose a modified dynamic programming method that uses two tables. One table stores the values of $D_c$ for sub-problems, as defined and proposed for \MTED~\cite{Zhang1996}. We introduce a  second table that stores $D'$, partial solutions to the modified edit distance for sub-problems. Figure~\ref{fig:origdptable} and Figure~\ref{fig:modifieddptable} show the dynamic programming table corresponding to $D_c$ and $D'$, respectively. Entries in the second table $D'$ are defined as follows:
\begin{enumerate}
    \item To compute the entry $(i,j)$, we need first populate the entries for the subtrees of $T_1[i]$ and $T_2[j]$. Say, we need to compute the entry at $(i_k,j_l)$, where $i_k \in T_1[i]$, $j_l \in T_2[j]$. If the subtrees rooted at $i_k$ and $j_l$ do not contain the unpaired nodes $i_u$ and $j_u$, \emph{i.e.,} $i_u \notin T_1[i_k]$ and $j_u \notin T_2[j_l]$,  then we pick the corresponding entry from $D_c$, namely $D_c (T_1[i_k], T_2[j_l])$. Figure~\ref{fig:origent} and Figure~\ref{fig:modent} shows the entries required to compute the entry $(i,j)$ in both tables. 
    \item If the subtrees rooted at $i_k$ and $j_l$ contain the unpaired nodes, we refer to the modified table $D'$, whose entries are computed using truncated persistence.
\end{enumerate}


Figure~\ref{fig:origdptable} shows the original table used to compute $D_c$. It does contain entries corresponding to different pairs of subtrees, but only the global entry $(i,j)$ corresponds to a distance between two merge trees. Other local comparisons involve subtrees that are not merge trees. Figure~\ref{fig:modifieddptable} is the modified table storing values of $D'$. Figures~\ref{fig:origent}  and~\ref{fig:modent} show the difference between the original and modified tables, and the entries that are required to calculate $D_c(i_{10},j_7)$ and $D'(i_{10},j_7)$. The labels correspond to the merge trees from Figure~\ref{fig:spt}. We denote tree distance entries by $d_c(,)$ and $d'(,)$, and forest entries by $f_c(,)$ and $f'(,)$. The node $i_5$ is unpaired in $T_1[i_{10}]$ and $j_4$ is unpaired in $T_2[j_7]$. In case of $D_c$, there is no distinction between paired and unpaired nodes. So, $D_c(i_{10},j_7)$ depends only on the following entries of the original table -- $d_c(i_{10},-1)$, $d_c(-1,j_7)$, $d_c(i_7,j_7)$, $d_c(i_8,j_7)$, $d_c(i_{10},j_5)$, $d_c(i_{10},j_6)$, and $f_c(i_{10},j_7)$.  In case of $D'$, for all subtrees involving unpaired nodes ($T_1[i_8],T_2[j_5]$), the corresponding entries are read from the modified table, while the remaining entries are read from the table for $D_c$. So, the required entries include $d'(i_{10},-1)$, $d'(-1,j_7)$, $d_c(i_7,j_7)$, $d'(i_8,j_7)$, $d'(i_{10},j_5)$, $d_c(i_{10},j_6)$, and $f'(i_{10},j_7)$. Further, the entries in the additional row and column are also required. They contain the truncated persistence values for the unpaired nodes corresponding to the current level.

\subsection{Algorithm and Analysis}
Zhang described an algorithm for computing the tree edit distance for labeled unordered trees~\cite{Zhang1996}. It is a dynamic programming based algorithm that follows from the properties discussed in Section~\ref{sec:constrainedmapping}. We adapt this  dynamic programming based method for computing \LMTED but compute and maintain two tables $D_c$ and $D'$ simultaneously. The algorithms fills entries in both tables $D_c$ and $D'$ iteratively. An entry within a table is computed and filled if entries corresponding to all sub-problems are already filled in previously. This implicitly corresponds to traversing the two input trees in a bottom up fashion in tandem. After both tables are filled, \LMTED is computed for all pairs of subtrees following the definition \emph{i.e.}, as a sum of $D'$ and $\Gamma$. In the worst case, the  algorithm computes \LMTED between all pairs of subtrees in $O(|T_1| \times |T_2| \times (deg(T_1)+deg(T_2)) \times log_{2}(deg(T_1)+deg(T_2)))$ time, similar to \MTED. Section 2 of the supplementary material describes the \LMTED algorithm in detail with pseudocode. In practice, the computation is restricted to a much smaller set of entries that need to be filled within both tables. This restricted set of entries is identified in a preprocessing step as described in the following section. Also, note that the time is amortized over all pairs of subtrees.  

%% file: section5.tex
\section{Refinement and Optimization}
\label{sec:refinementoptimization}
Given two merge trees, we observe that in many applications it is unnecessary to compute \LMTED between all pairs of subtrees. This section describes refinements that reduces the number of pairs of subtrees that are considered for \LMTED computation. This optimization leads to faster computation times. This step may be skipped if it is necessary to compare all pairs of subtrees. We also describe a refinement that ensures that all subtrees considered for comparison are merge trees.

\subsection{Ordering subtrees}
\label{sec:ordersubtrees}
The dynamic programming algorithm works for any ordering of nodes. The entries required to compute the distance between a pair of subtrees are computed if they are not already available from the table. However, we choose to order the nodes by assigning a priority based on the size of the subtree rooted at the node and on the number of grid points in the domain mapped to the subtree (\emph{i.e.}, volume of the corresponding region in the domain). This ordering facilitates easy identification of similar regions via visual inspection of the distance matrix because subtrees of similar size appear in the close vicinity of one another within the matrix.

\subsection{Comparison refinement}
Since \LMTED is computed between all pairs of subtrees of the two input trees, the number of comparisons is determined by the size of the two trees. In general, the scalar fields to be compared are unrelated, defined on different domains, and have different ranges. So, comparing all pairs of subtrees is necessary and unavoidable. However, in several applications, it is not necessary or meaningful to compare all pairs of subtrees. We describe two such scenarios to motivate a refinement step that reduces the number of comparisons.
\begin{enumerate}
\item \textbf{Symmetry detection:} The scalar field is compared with itself. So, we can discard comparisons between subtrees with vastly different sizes (for example, $T_1[i_4]$ and $T_1[i_{10}]$ in Figure~\ref{fig:spt}) or a subtree that is contained within another subtree (like $T_1[i_{11}]$ and $T_1[i_8]$).
\item \textbf{Time-varying data analysis:} While analyzing a time-varying scalar field, assuming a fine enough temporal resolution, we may discard comparisons between subtrees with vastly different sizes. We may also discard comparisons between subtrees that map to regions in the domain with significantly different sizes (area / volume). 
\end{enumerate}

We define a set of criteria used to direct the refinement. Each criterion is a ratio between measures or a statistic computed for a subtree. Let $T_1,T_2$ be the two input merge trees with nodes $\{i_1,i_2, \ldots, i_{n_i}\}$ and $\{j_1,j_2, \ldots, j_{n_j}\}$, and roots $r_1$ and $r_2$, respectively. 
Consider a subtree  $T_1[i_k]$ and the mapping to its associated region in the domain. This region is a connected component of the preimage of the range of scalar values corresponding to $T_1[i_k]$. Assuming that the scalar function is defined over a 3D domain, the volume of this region may be approximated by counting the number of sample points (vertices or grid points, depending on whether the domain is represented using a tetrahedral mesh or a cube grid). 
The aggregate persistence $P_{i_k}$ of the subtree is computed as the sum of persistence of all persistence pairs contained within the subtree. Given a pair of subtrees $T_1[i_k]$ and $T_2[j_l]$, we use the ratio between
\begin{enumerate}
    \item number of nodes in the subtrees $\left| T_1[i_k] \right| / \left| T_2[j_l] \right|$,  
    \item volume (or area) of the domain that maps to the two subtrees, and 
    \item aggregate persistence of the subtrees $ P_{i_k} / P_{j_l}$
\end{enumerate}
to determine the refinement. Thresholds for the ratios are determined empirically. For each criterion, we plot the number of pairs of subtrees against increasing values of the ratio, identify the value of the ratio corresponding to a sharp decline or the 'knee' of the curve, and choose this value of the ratio as threshold. If the plot does not exhibit a clear knee then we set the threshold to  $0.5$. Further, if $T_1 = T_2$ then we discard all comparisons between $T_1[i_k]$ and all subtrees contained within $T_1[i_k]$.

\subsection{Subtree refinement}
Next, we preprocess the input to ensure that we compare only those subtrees that are merge trees. 

 The subtrees that constitute the sub-problems in the dynamic programming algorithm for \MTED~\cite{Sridh2020} are not necessarily merge trees. Consider the scenario in Figure~\ref{fig:spt}. While trees $T_1[i]$ and $T_2[j]$ rooted at $i$ and $j$ are merge trees, $T_1[i_8]$ and $T_2[j_5]$ are not merge trees as they contain unpaired nodes, namely $i_{5}$ and $j_4$.  The entry $D_c(i_8,j_5)$ in the DP table  (Figure~\ref{fig:origdptable}) contains a value that is used to computed the \MTED between $T_1[i]$ and $T_2[j]$. But, it is not a meaningful distance between subtrees $T_1[i_8]$ and $T_2[j_5]$ because the cost model used for operations related to unpaired nodes $i_5,j_4$ depends on their original persistence. From Figure~\ref{fig:local}, it is clear that the regions associated to the two subtrees are similar to each other but the value of $D_c(i_8,j_5)$ does not reflect this similarity. 

We insert a dummy node $i'_{10}$ in $T_1[i_8]$ with  $|f_1(i'_{10})-f_1(i_{10})| < \varepsilon$ for a small $\varepsilon >0$. In subsequent computations, we consider $T_1[i'_{10}]$ as the merge tree corresponding to the subtree $T_1[i_8]$. Similarly, $T_2[j_5]$ contains the unpaired node $j_4$. We insert a dummy node $j'_7$ and  consider $T_2[j'_7]$ as the merge tree corresponding to $T_2[j_5]$.  This conversion is consistent with the mapping between subtrees and regions of the domain and hence results in meaningful distances. 

%% file: section6.tex
\section{Applications}
In this section, we demonstrate the utility of \LMTED in applications like symmetry detection, feature tracking, and spatio-temporal exploration of scientific data. We also describe results of a comprehensive analysis of the effects of subsampling, smoothing, and topologically controlled compression. In all cases where a global comparison is meaningful, results based on \MTED~\cite{Sridh2020} are taken as a baseline.
\subsection{Understanding the local tree edit distance}
\begin{figure}
    \centering
    \includegraphics[width=0.25\textwidth]{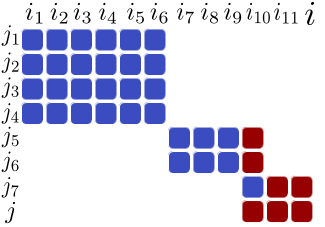}
    \caption{Understanding \LMTED. Columns of the distance matrix (DM) represent subtrees rooted at nodes of merge tree $T_1$,  rows represent subtrees rooted at nodes of tree $T_2$. Nodes are ordered as per the priority described in Section~\ref{sec:ordersubtrees}. \LMTED values are shown using a blue-red colormap ($0$~\protect\includegraphics[height=0.15cm]{Images/legend2.png}~$0.1$)}.
    \label{fig:uloc}
    \vspace{-0.15in}
\end{figure}
We begin with a simple study to understand \LMTED, by comparing two scalar fields shown in Figures~\ref{fig:local}, whose split trees are shown in Figure ~\ref{fig:spt}. Figure~\ref{fig:uloc} shows the distance matrix (DM), entries corresponding to subtree pairs that are discarded during the refinement step are blank. We observe two blue blocks of size $4 \times 6$ and $2 \times 3$ in the DM, confirming that there are two sets of similar regions and the pair of similar subtrees $T_1[i_{10}],T_2[j_7]$. Note that the distances between these similar regions are very small $ \le 0.000093$ in contrast to the larger value of \MTED ($=0.33$) between the two trees. Such instances of local similarity without global similarity is common in scientific data. Further, \LMTED also captures similarity at different scales.

\subsection{Symmetry Detection}
\begin{figure}
    \centering
    \subfigure[Volume rendering of the Rubisco RbcL8-RbcX2-8 complex(EMDB-1654)]{\includegraphics[width=0.39\textwidth]{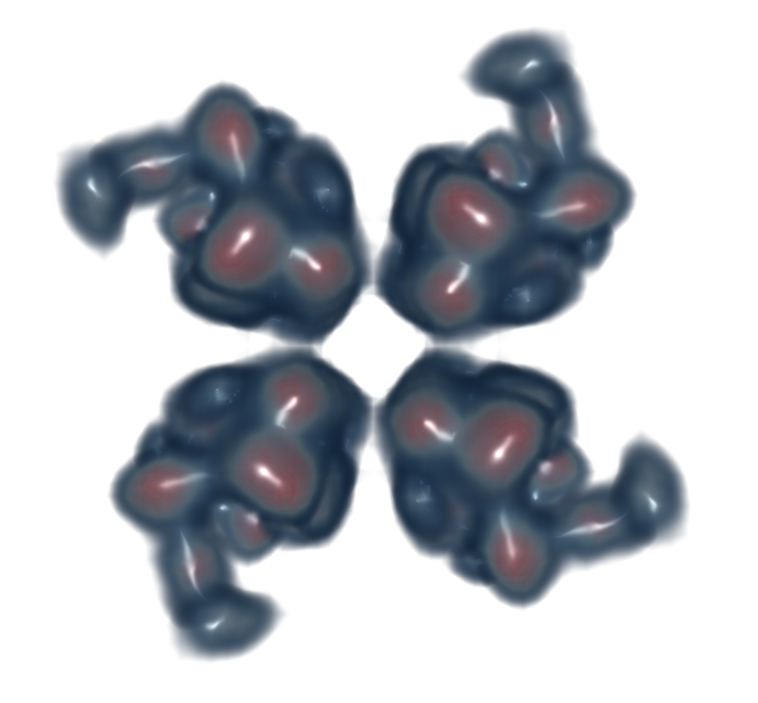}\label{fig:vol1654}}
    ~
    \subfigure[Distance Matrix (DM)]{\includegraphics[width=0.43\textwidth]{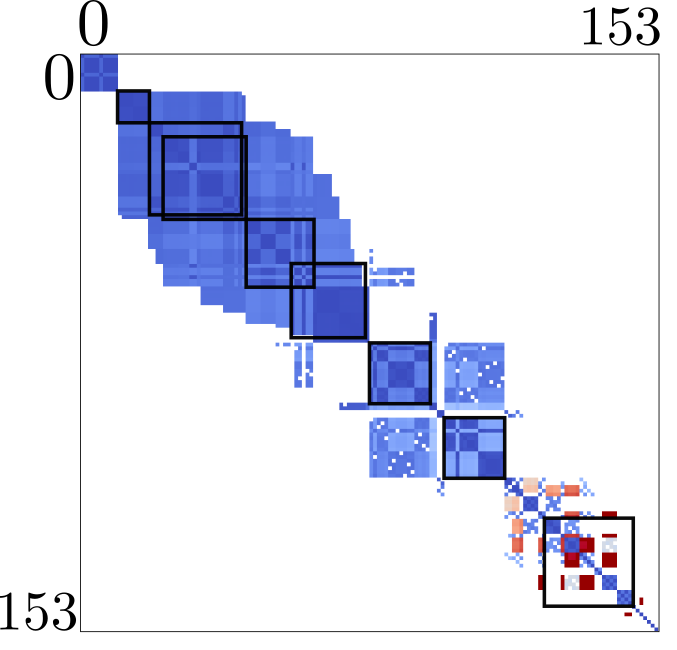}\label{fig:DM1654}}
    \caption{\LMTED values in the DM are shown using a blue-red colormap ($0$~\protect\includegraphics[height=0.15cm]{Images/legend2.png}~$0.1$)}.
    \label{fig:1654}
    \vspace{-0.15in}
\end{figure}
Finding symmetric structures in scalar fields is a challenging problem~\cite{thomas2011,thomas2013,thomas2014}). The \MTED driven approach~\cite{Sridh2020} extracts a particular set of high persistent subtrees that are known to be symmetric and compares them to verify symmetry. We take a different approach where we detect symmetry directly based on local similarity by comparing the scalar field with itself. We use CryoEM data from EMDB~\cite{EMDB2021}, which contains 3D electron microscopy density data of macromolecules, subcellular structures, and viruses. We first compute the simplified merge tree (using a small persistence threshold $<1\%$) and consider pairs of subtrees after refinements described in Section~\ref{sec:refinementoptimization}. 

\begin{figure}
    \centering
\subfigure[DM $124$]{\includegraphics[width=0.22\textwidth]{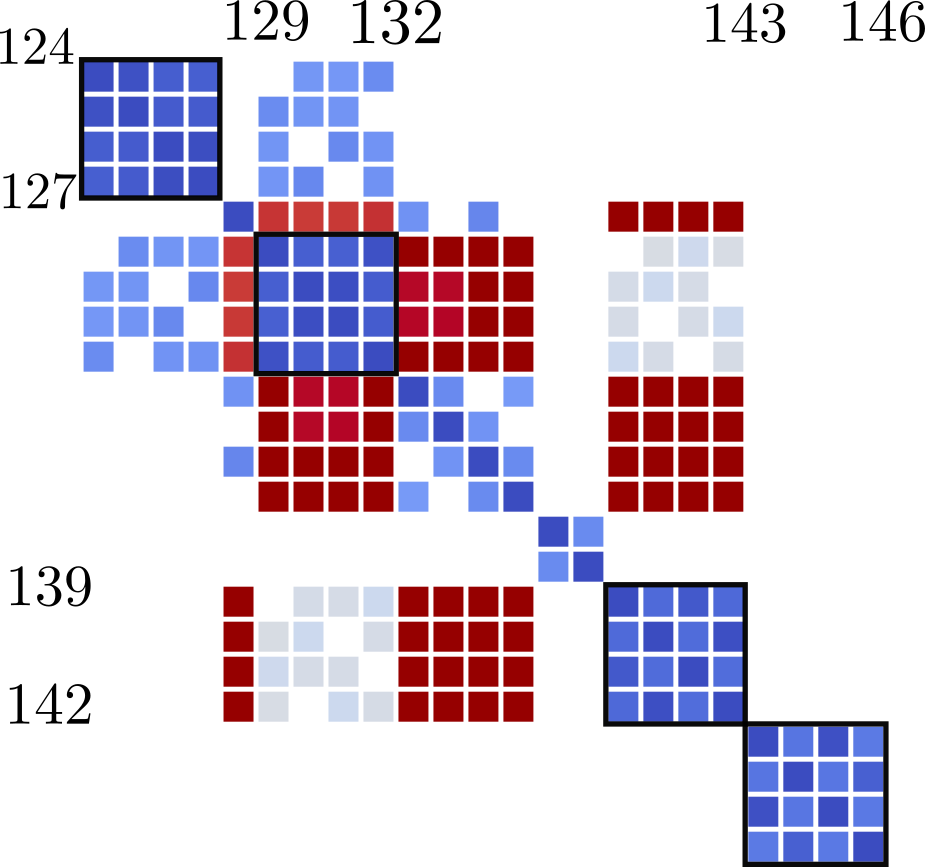}
\label{fig:mat124}}
~
\subfigure[DM $97$]{\includegraphics[width=0.22\textwidth]{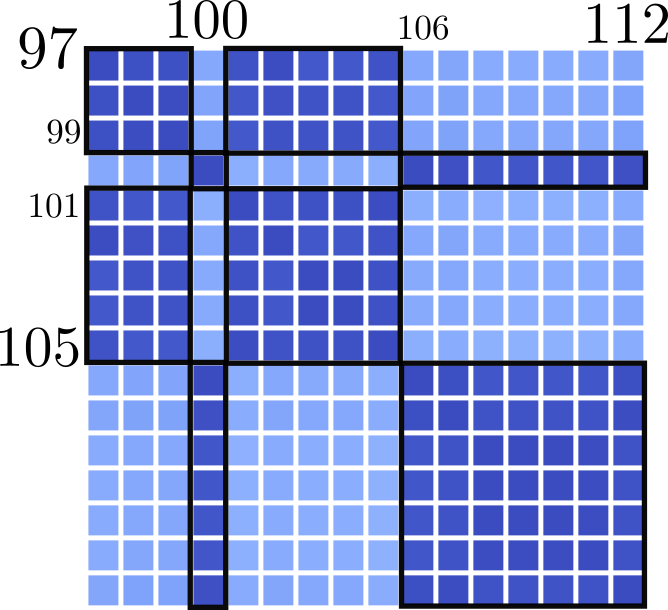}
\label{fig:mat97}}

\subfigure[volume $97$]{\includegraphics[width=0.14\textwidth]{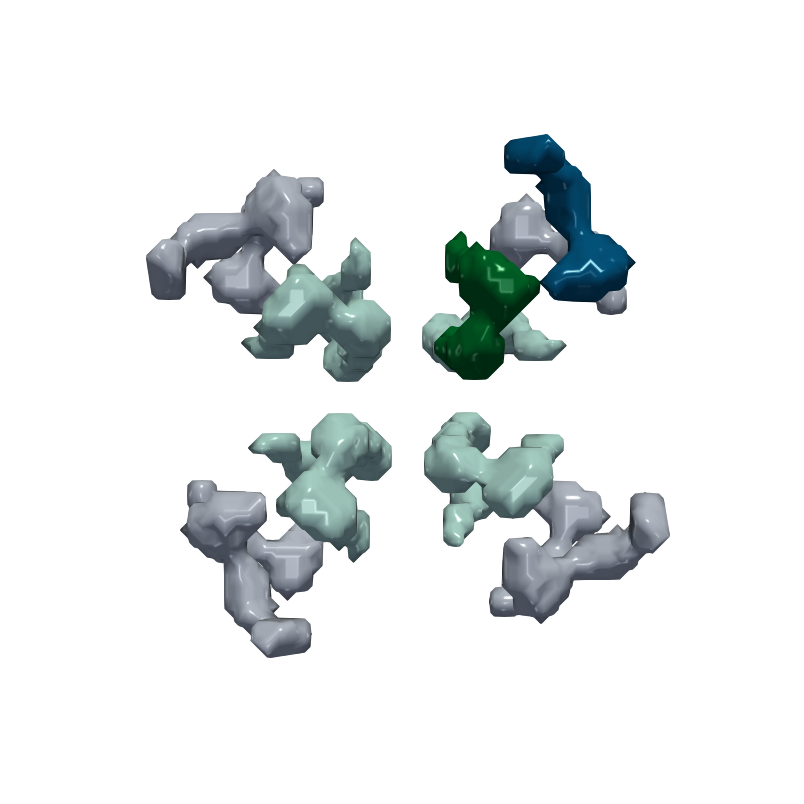}
\label{fig:vol97}}
~
\subfigure[volume $124$]{\includegraphics[width=0.14\textwidth]{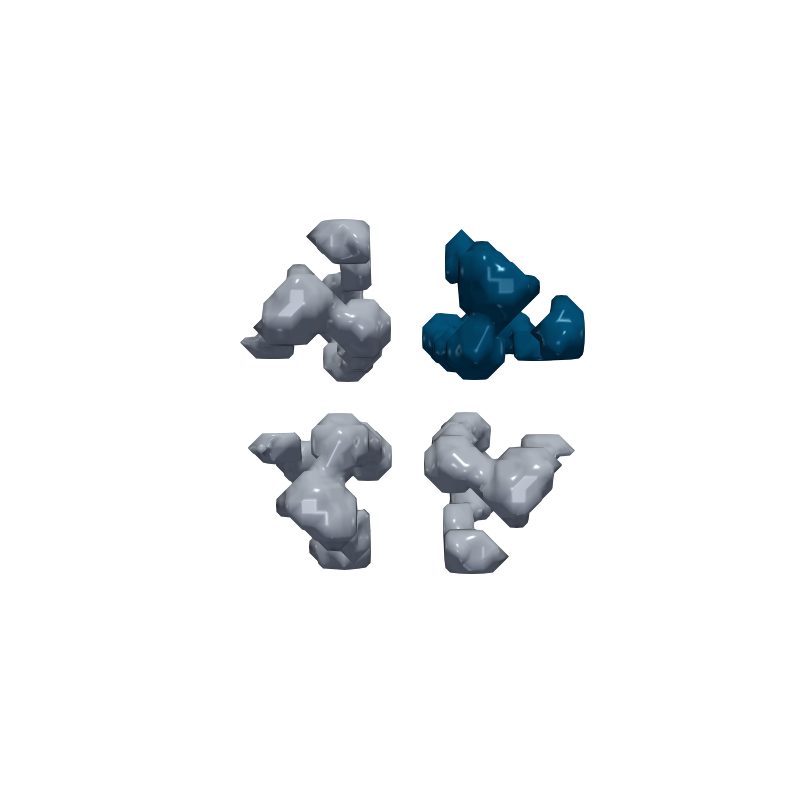}
\label{fig:vol124}}
~
\subfigure[volume $129$]{\includegraphics[width=0.14\textwidth]{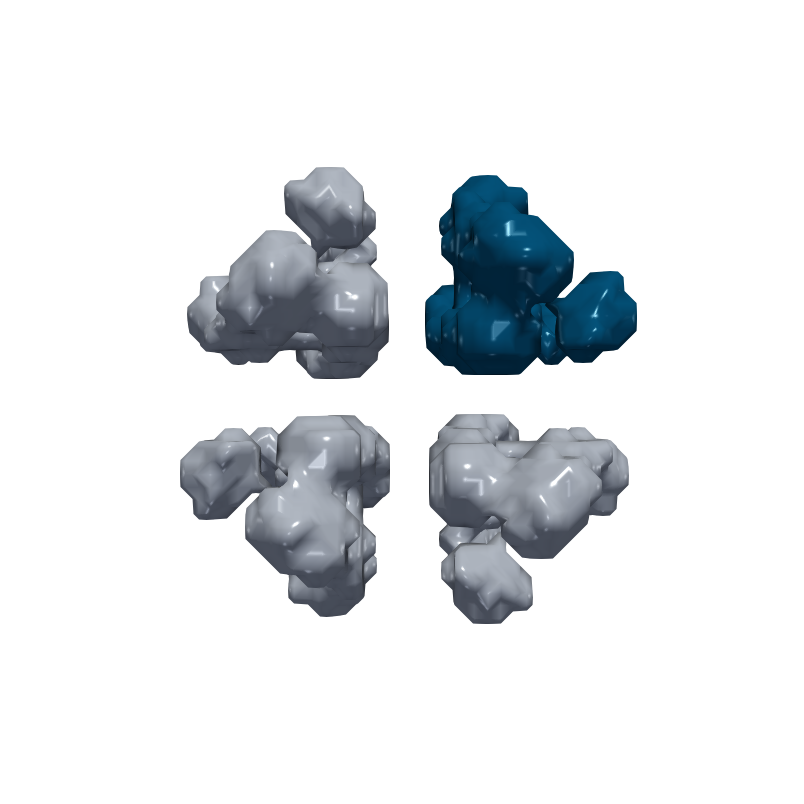}
\label{fig:vol129}}

\subfigure[volume $139$]{\includegraphics[width=0.22\textwidth]{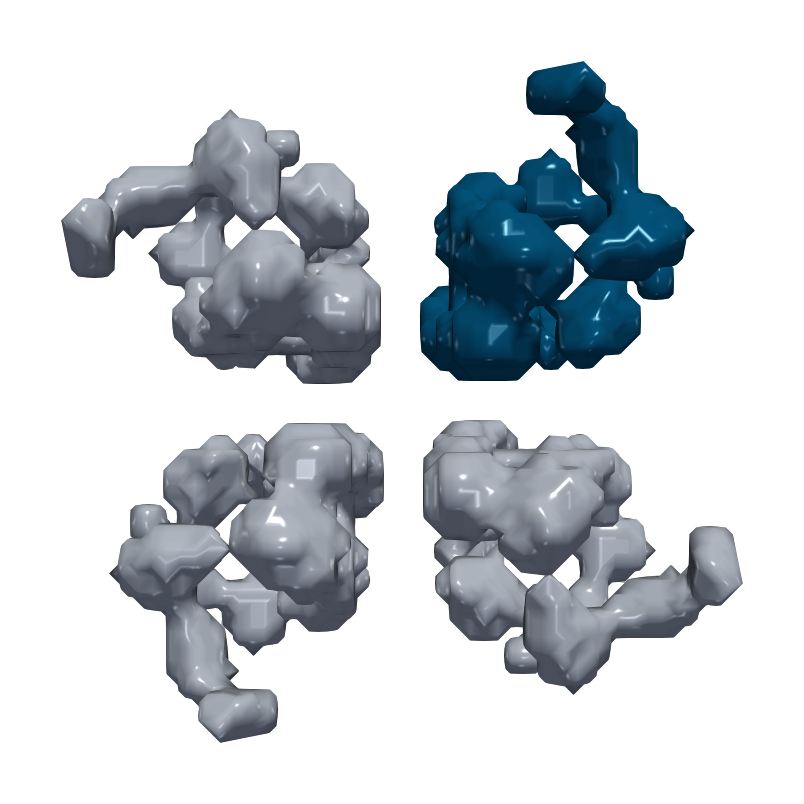}
\label{fig:vol139}}
~
\subfigure[volume $143$]{\includegraphics[width=0.22\textwidth]{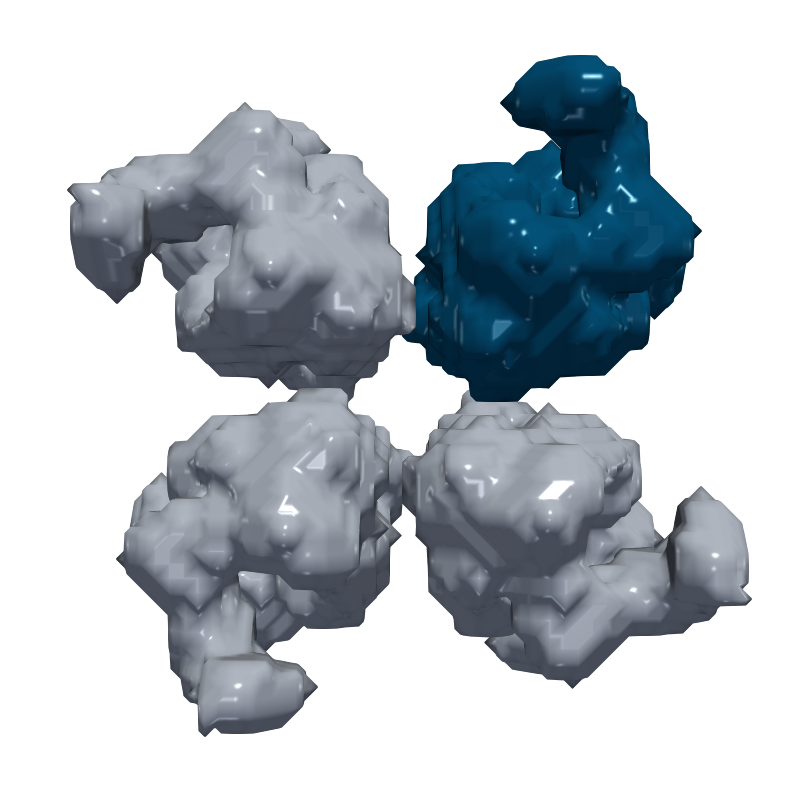}
\label{fig:vol143}}

\caption{Highlighted submatrices from Figure~\ref{fig:1654} and corresponding regions. \subref{fig:vol124}-\subref{fig:vol143}~Regions corresponding to submatrices highlighted in~\subref{fig:mat124}. \subref{fig:vol97}~Region corresponding to submatrix shown in~\subref{fig:mat97}}
\label{fig:allregions1654}
\vspace{-0.15in}
\end{figure}

\begin{figure}
    \centering
\subfigure[DM $10$]{\includegraphics[width=0.11\textwidth]{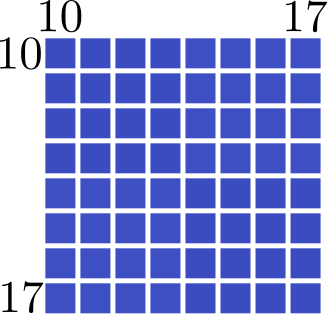}
\label{fig:mat10}}
~
\subfigure[volume $10$]{\includegraphics[width=0.1\textwidth]{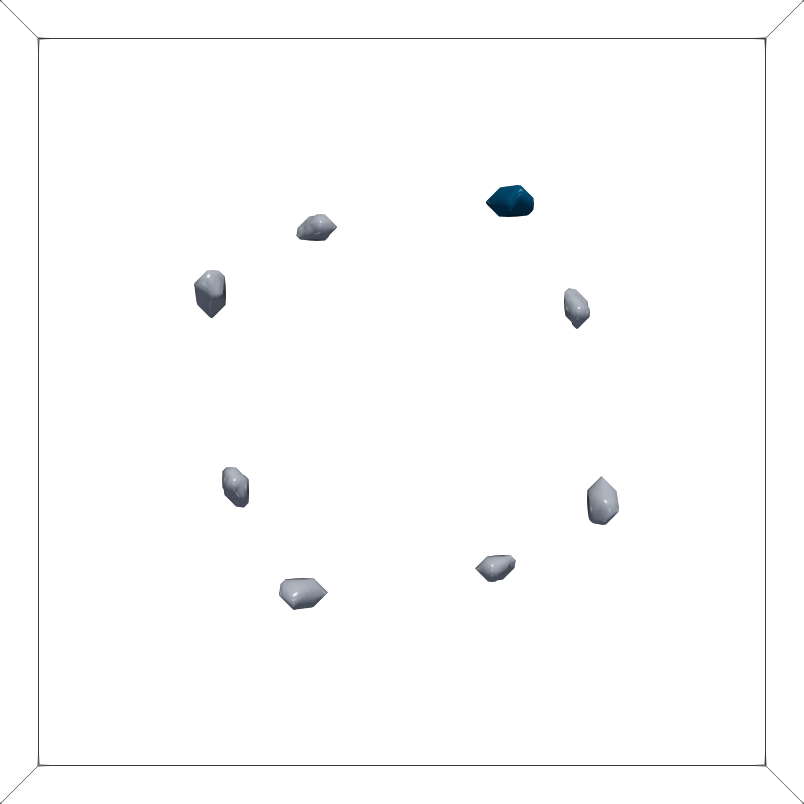}
\label{fig:vol10}}
~
\subfigure[DM $18$]{\includegraphics[width=0.11\textwidth]{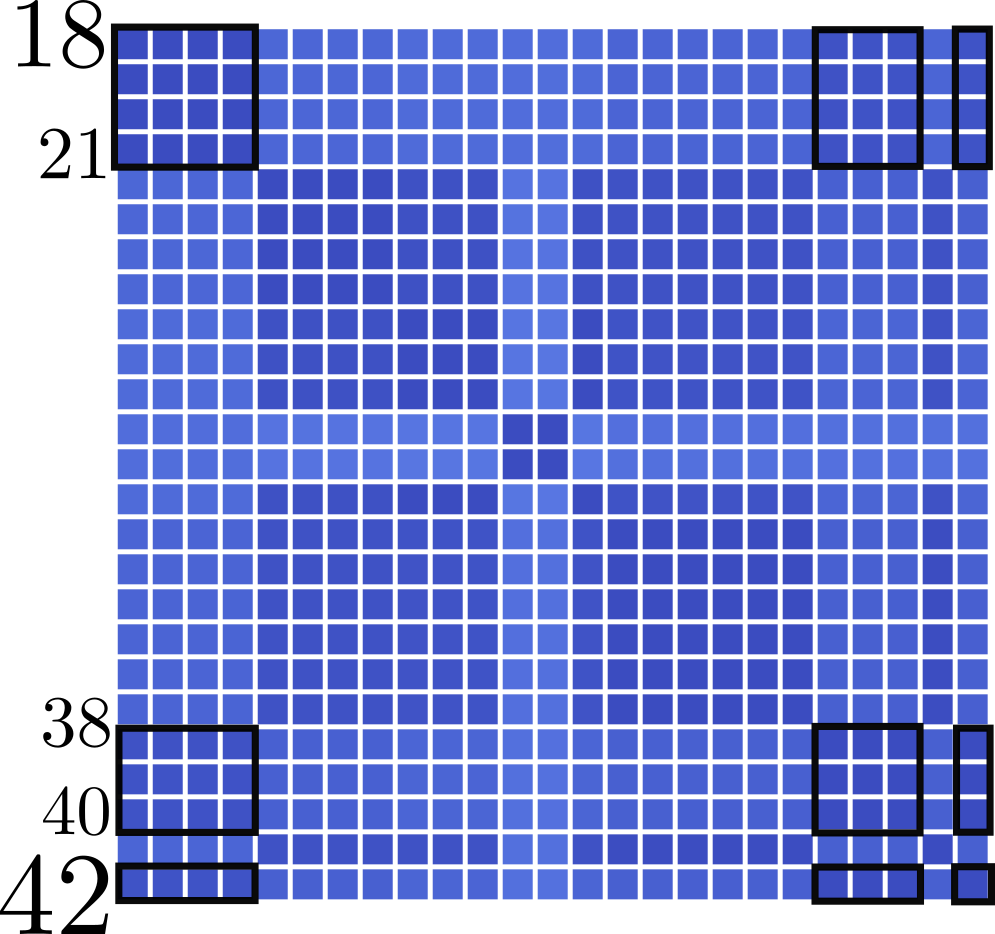}
\label{fig:mat18}}
~
\subfigure[volume $18$]{\includegraphics[width=0.1\textwidth]{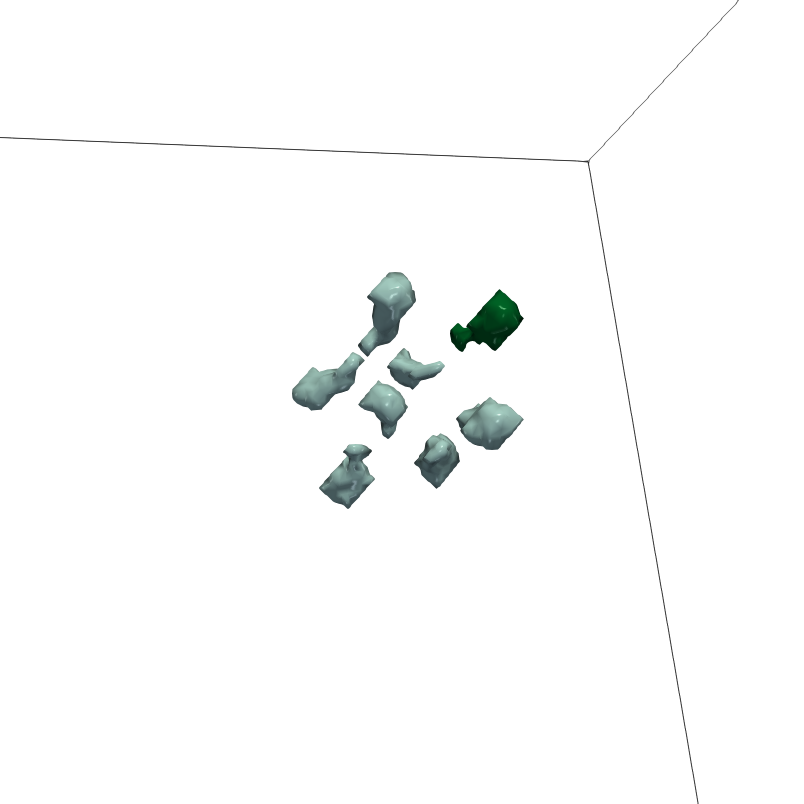}
\label{fig:vol18}}

\subfigure[DM $22$]{\includegraphics[width=0.11\textwidth]{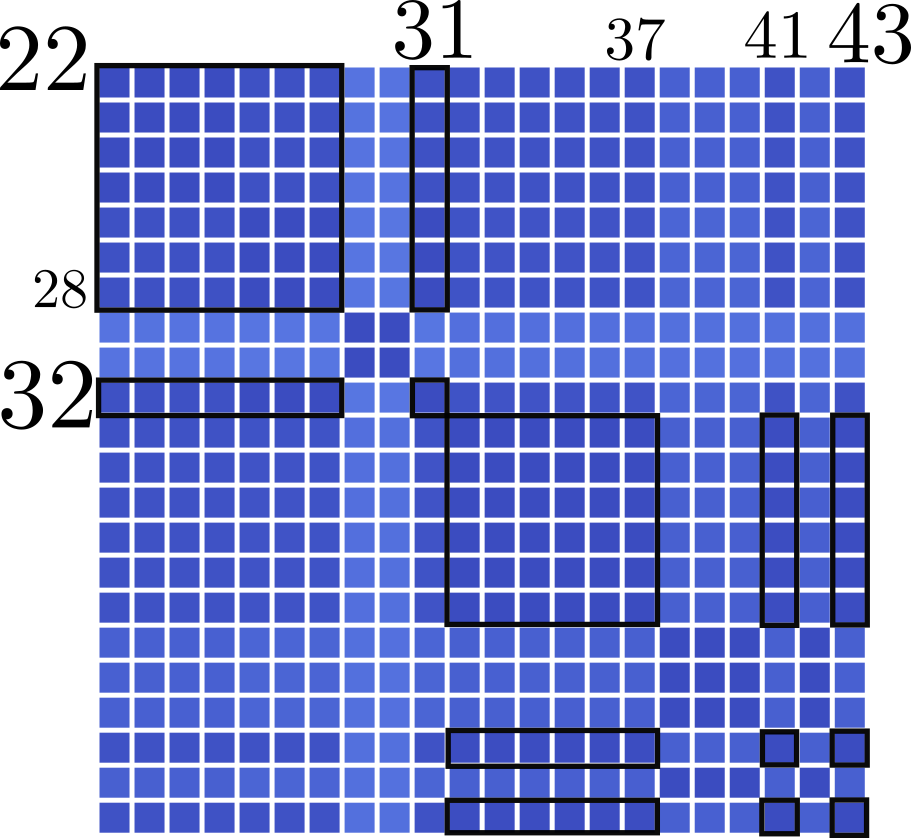}
\label{fig:mat22}}
~
\subfigure[volume $22$]{\includegraphics[width=0.1\textwidth]{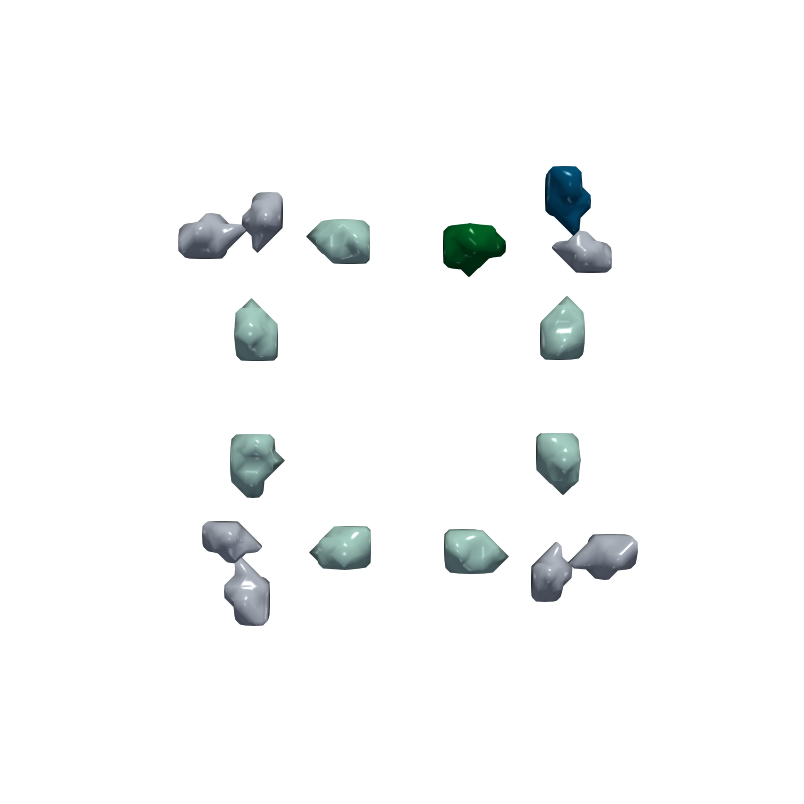}
\label{fig:vol22}}
~
\subfigure[DM $44$]{\includegraphics[width=0.11\textwidth]{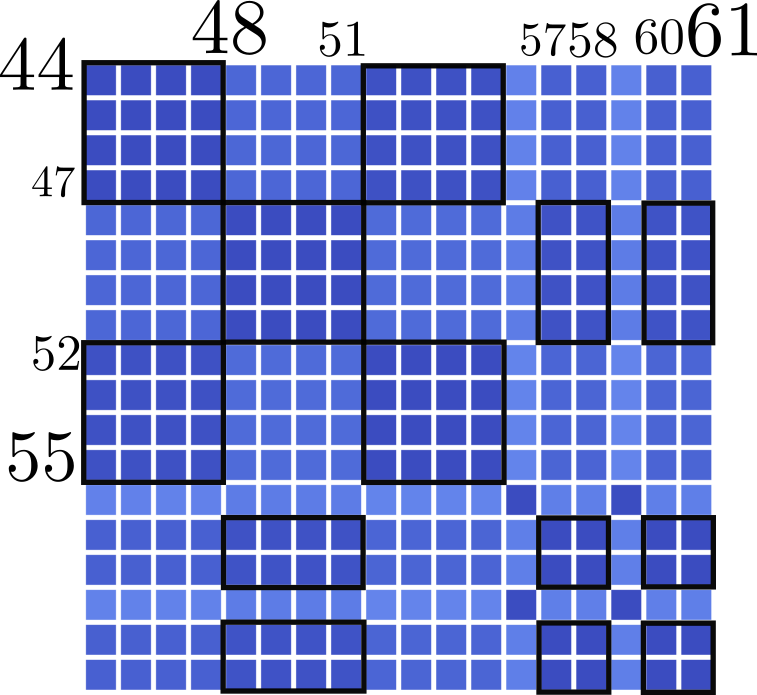}
\label{fig:mat44}}
~
\subfigure[volume $44$]{\includegraphics[width=0.1\textwidth]{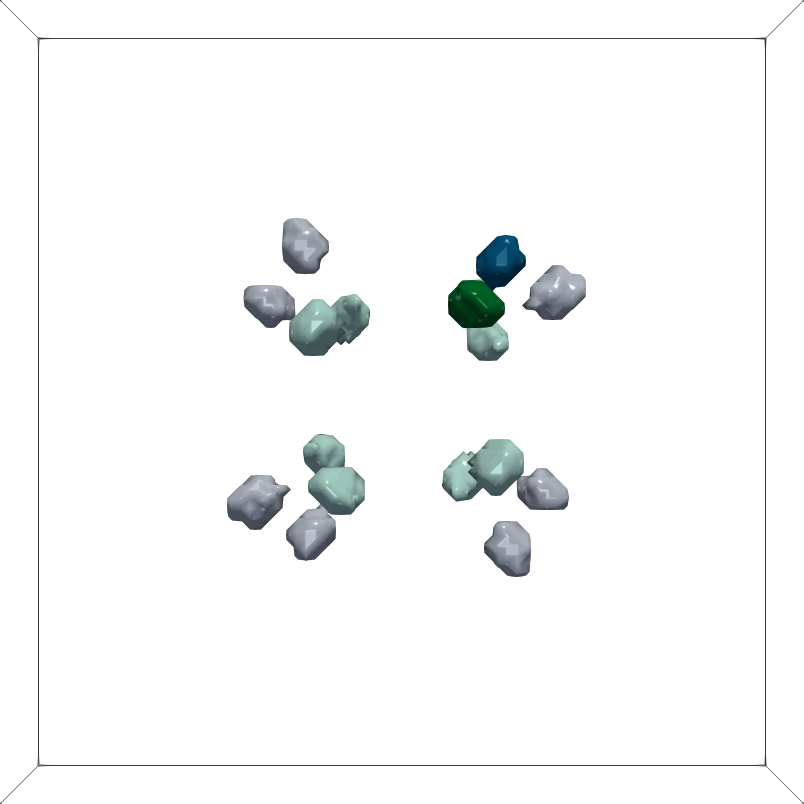}
\label{fig:vol44}}

\subfigure[DM $56$]{\includegraphics[width=0.11\textwidth]{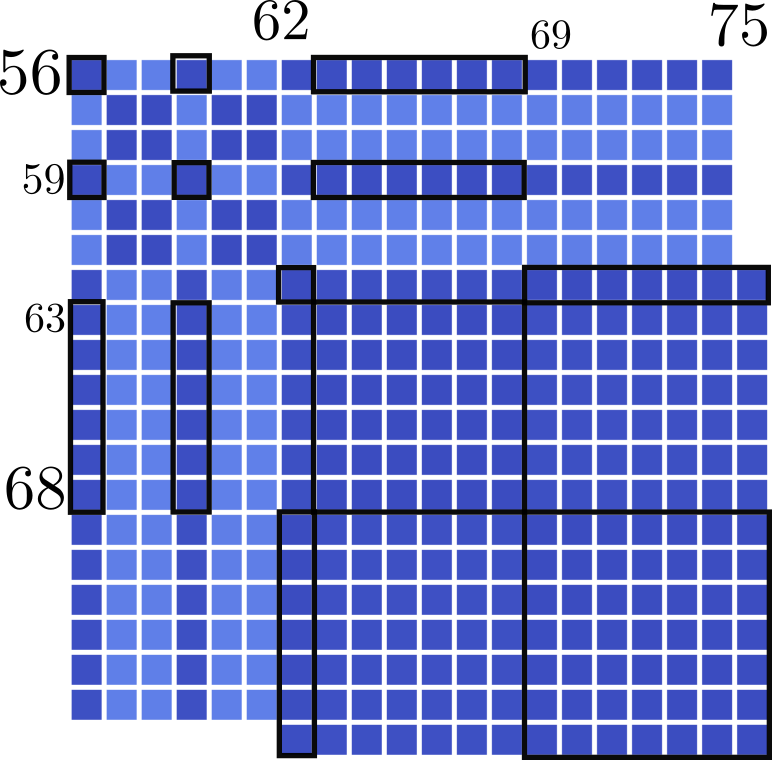}
\label{fig:mat56}}
~
\subfigure[volume $56$]{\includegraphics[width=0.1\textwidth]{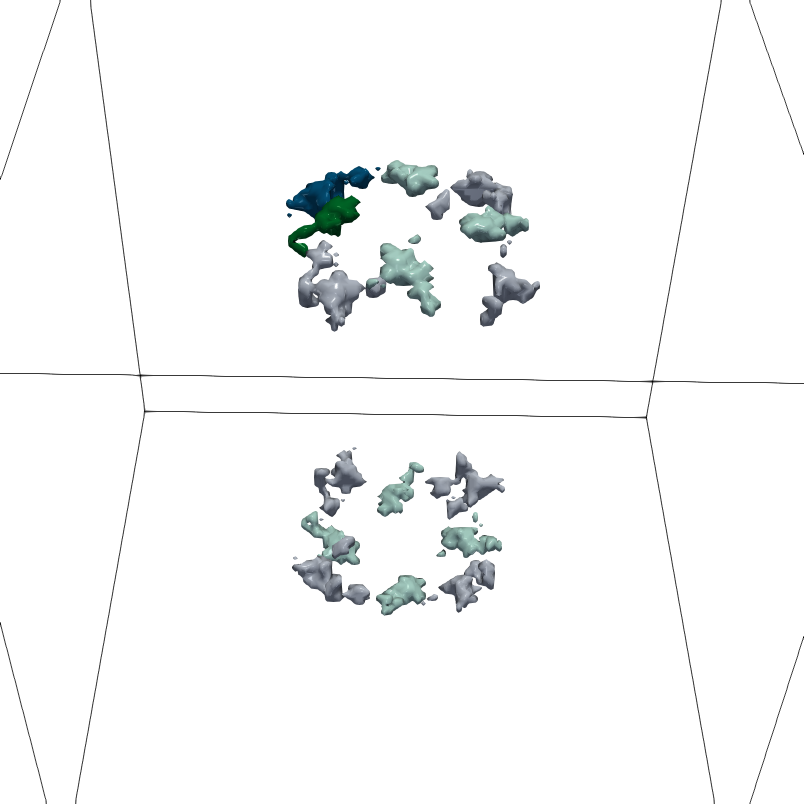}
\label{fig:vol56}}
~
\subfigure[DM $77$]{\includegraphics[width=0.11\textwidth]{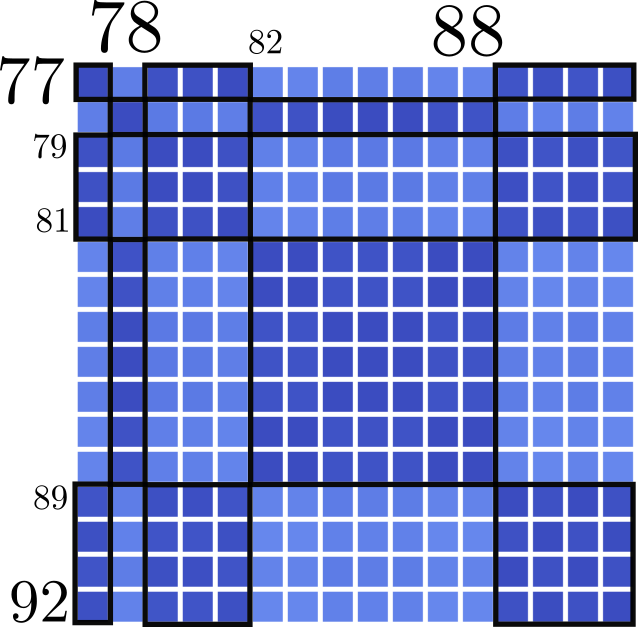}
\label{fig:mat77}}
~
\subfigure[volume $77$]{\includegraphics[width=0.1\textwidth]{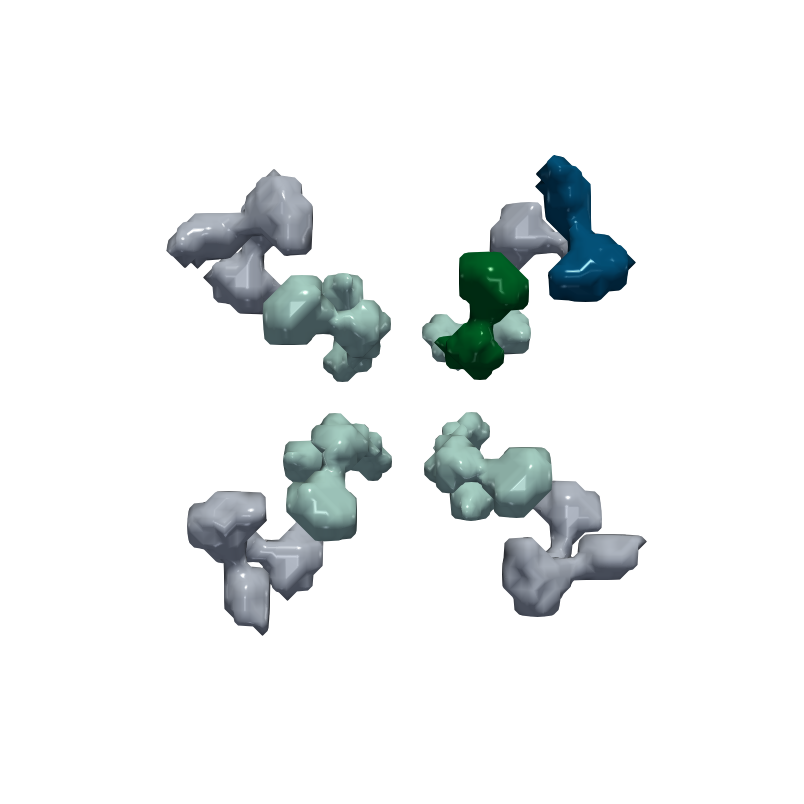}
\label{fig:vol77}}

\caption{Smaller regions of the Rubisco RbcL8-RbcX2-8 complex (EMDB-1654) corresponding to the highlighted submatrices.}
\label{fig:regions1654}
\vspace{-0.15in}
\end{figure}

We illustrate and analyse the results using the Rubisco RbcL8-RbcX2-8 complex (EMDB-1654) shown in Figure~\ref{fig:vol1654}. We compute \LMTED between all pairs of subtrees of its merge tree after the refinement step. The resulting  DM is shown in Figure~\ref{fig:DM1654}. Blank regions correspond to subtree pairs that are discarded during the refinement step. Submatrices highlighted in black correspond to regions in the data that are symmetric. For clarity, we have shown these submatrices together with the corresponding regions in Figures~\ref{fig:allregions1654},~\ref{fig:regions1654}. We observe that \LMTED detects symmetric regions at different scales.

Any selection of submatrices from Figure~\ref{fig:DM1654} with a common color corresponds to a set of symmetric regions, we highlight some of them. In some cases matrix entries corresponding to symmetric regions may not appear adjacent to each other as a submatrix. But, it may be possible to visually identify the entries as belonging to a single cluster. A row/column reordering helps the identification of these clusters, see Behrisch et al.~\cite{behrisch2016} for details. The reordering may be restricted to a chosen submatrix to save computation time. We describe a few additional experimental results together with row/column reordering in the supplementary material.

\myparagraph{Comparison with previous methods.}
Thomas and Natarajan~\cite{thomas2011} process the branch decomposition of contour trees by building feature descriptors, and use them to identify similar subtrees. The main limitation of this approach to symmetry detection is that it is based exclusively on the structure and may fail when symmetric regions do not manifest as repeating subtrees. For example, if the field is noisy, subtrees corresponding to noise have high persistence, or when the field has large flat regions. Their proposed hierarchy descriptor and similarity measure is a good estimate but not as accurate as examining the complete hierarchy. It also ignores the geometry of repeating regions leading to regions with different geometry grouped together and regions with similar geometry grouped differently. We use grid points mapped to subtrees, as an easy-to-compute substitute for geometric information. This also helps us to find symmetry in multiple scales. We use merge trees instead of contour trees and avoid computation of extremum graphs, geodesic distances, or contour shape descriptors in contrast to previous methods~\cite{thomas2013,thomas2014}.  \LMTED computation is costly compared to the hierarchy descriptor based comparison~\cite{thomas2011}. We observe results similar to previous methods based on explicit geometric shape descriptors~\cite{thomas2014}, but a theoretical guarantee requires further study.

\subsection{Analysis of subsampling, smoothing, and topology based compression}
We analyse the effects of subsampling, smoothing and topology based compression~\cite{Soler2018a}. While subsampling and smoothing is applied uniformly across the domain, the effects of compression vary in different parts of the domain. We showcase how \LMTED can be used to analyse these effects meaningfully.

\begin{figure}
\centering
\vspace{-0.05in}
\subfigure[scalar function $f_2$]{\includegraphics[width=0.14\textwidth]{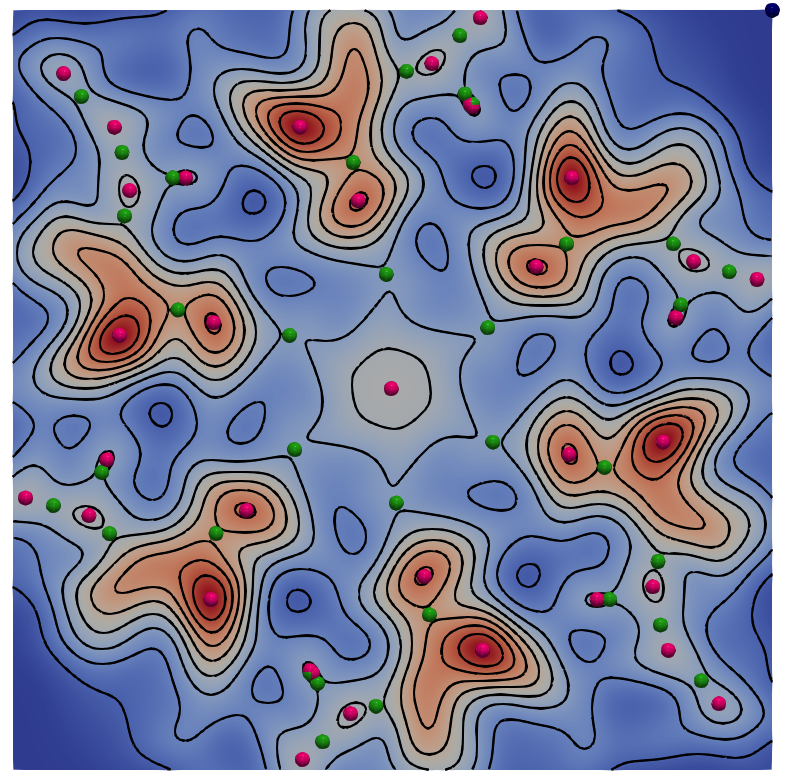}
\label{fig:ss4}}
~
\subfigure[subsampled $f_2$]{\includegraphics[width=0.14\textwidth]{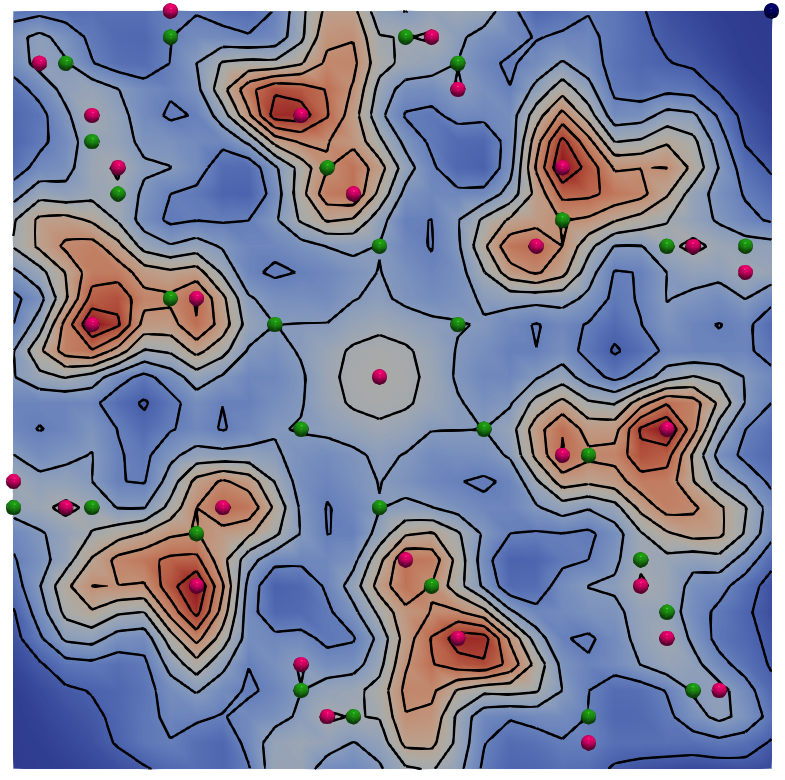}
\label{fig:ss5}}
~
\subfigure[smoothened $f_2$]{\includegraphics[width=0.14\textwidth]{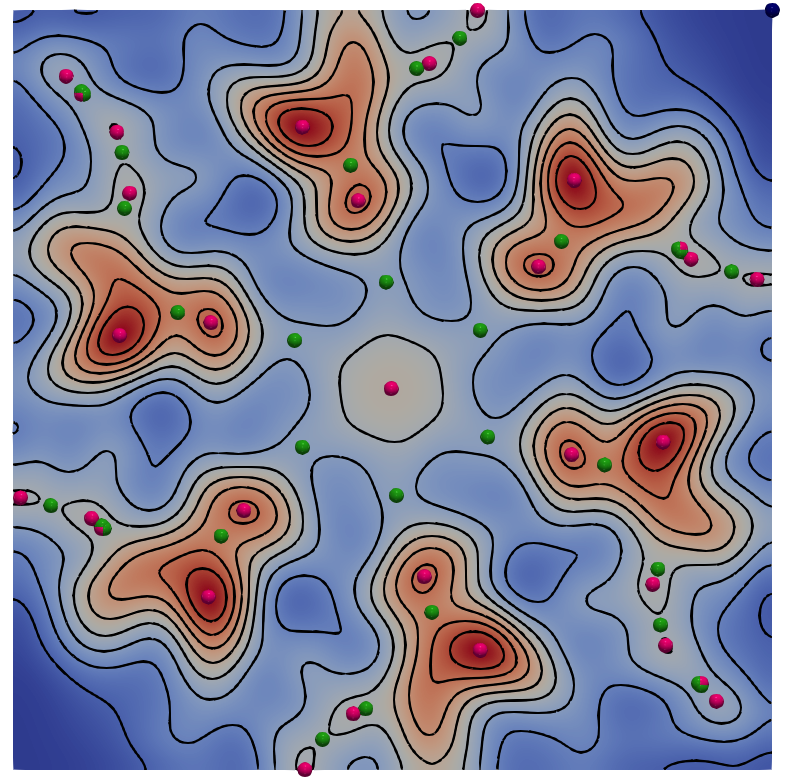}
\label{fig:ss6}}

\subfigure[DM for $f_2$, original and subsampled]{\includegraphics[width=0.22\textwidth]{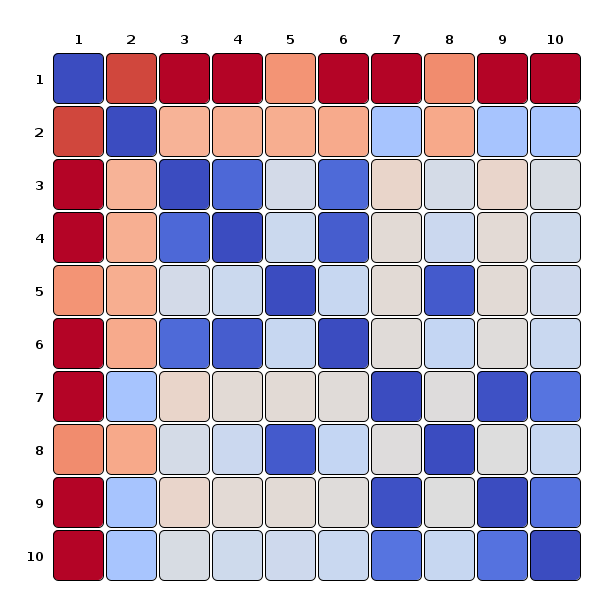}
\label{fig:sub2}}
~
\subfigure[DM for $f_2$, original and smoothened]{\includegraphics[width=0.22\textwidth]{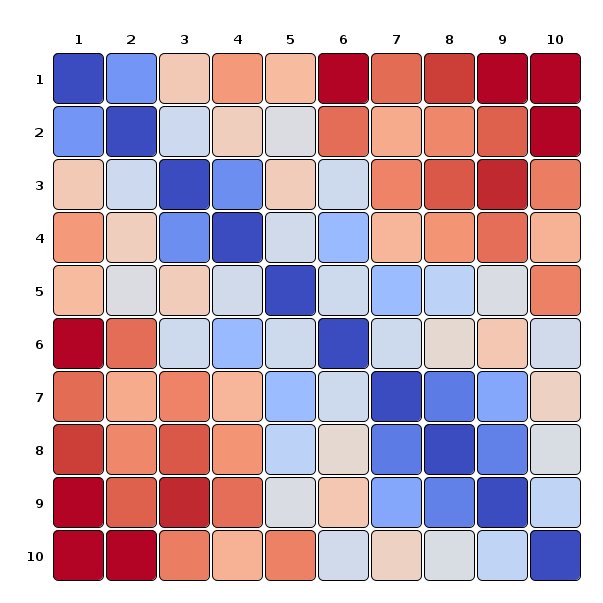}
\label{fig:smo2}}

\caption{Measuring the effect of subsampling and smoothing (Images sourced from Figure 14 from~\cite[Section 5.4]{Sridh2020}). \subref{fig:ss4}~A synthetic function $f_2$ sampled over a $300 \times 300$ grid.
\subref{fig:ss5}~$f_2$ subsampled down to a $30 \times 30$ grid over $9$ iterations.
\subref{fig:ss6}~$f_2$ smoothed in $9$ iterations.
\subref{fig:sub2}~DM showing distance between all pairs of subsampled datasets. 
\subref{fig:smo2}~DMs showing distances between all pairs of smoothed functions. Row and column indices correspond to the iteration number, $1$ corresponds to the lowest resolution/extreme smoothing, $10$ corresponds to the original. We again use a blue-red colormap (low~\protect\includegraphics[height=0.15cm]{Images/legend2.png}~high). The scales on colormaps for \subref{fig:sub2} and \subref{fig:smo2} are different
}
\label{fig:ss}
\vspace{-0.25in}
\end{figure}

\myparagraph{Effects of subsampling and smoothing. }
Topology changes due to subsampling and smoothing are not thoroughly quantified. While previous work does present some analysis based on the \MTED, it is global and  not capable of providing fine-grained analysis or explain the non-monotonic variation in many cases. We present a fine-grained analysis using \LMTED on a scalar field denoted as $f_2$~\cite[Section 5.4]{Sridh2020} and we use the images of the scalar fields and the DMs from ~\cite[Figure 14]{Sridh2020} in Figure~\ref{fig:ss} to illustrate the benefits.

\begin{figure}
\centering

\subfigure[\LMTED DM for subtree pair  $(T_3,T_4)$, size $62 \times 66$]{\includegraphics[width=0.21\textwidth]{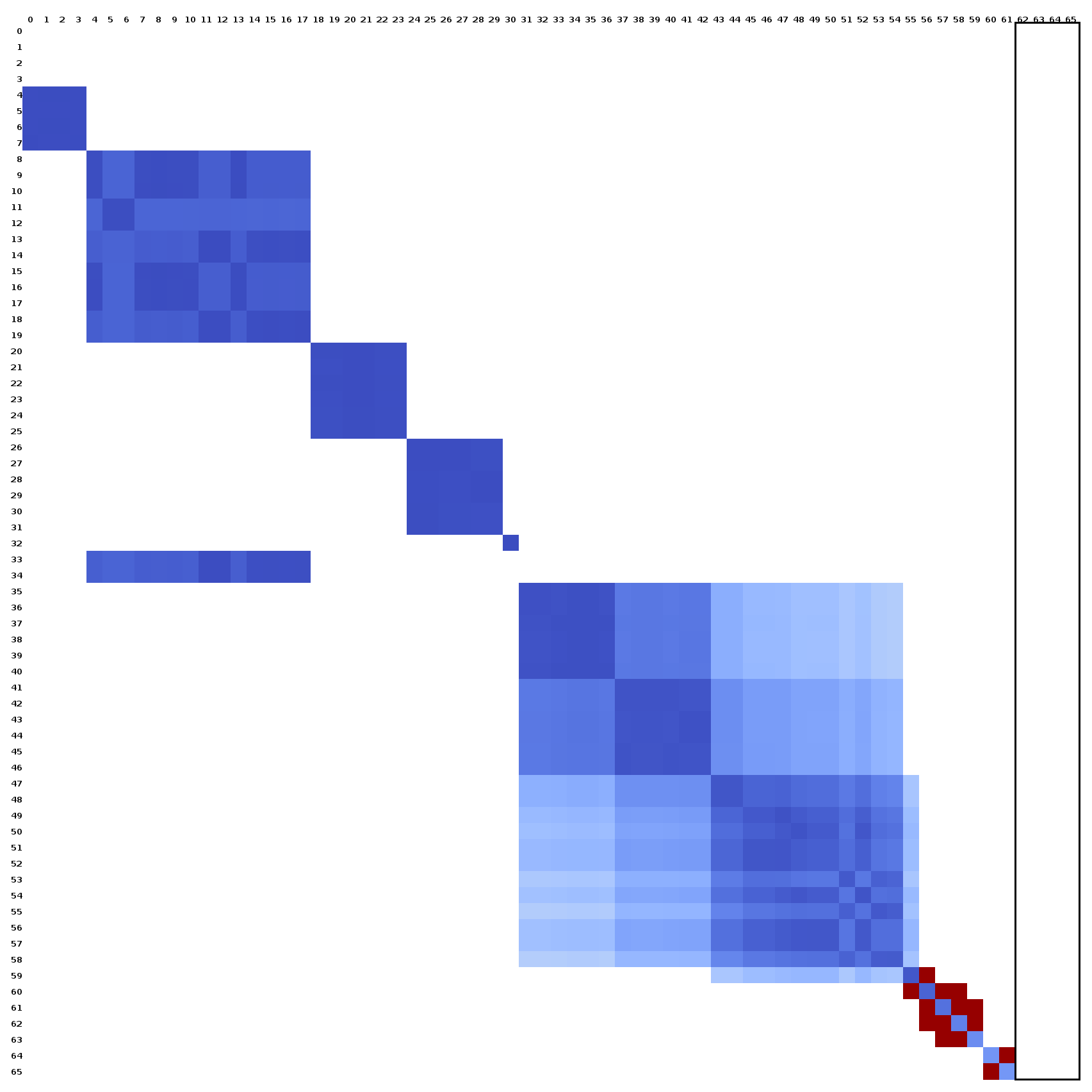}
\label{fig:sub34}}
~
\subfigure[\LMTED DM for subtree pair  $(T_3,T_5)$, size $62 \times 62$]{\includegraphics[width=0.21\textwidth]{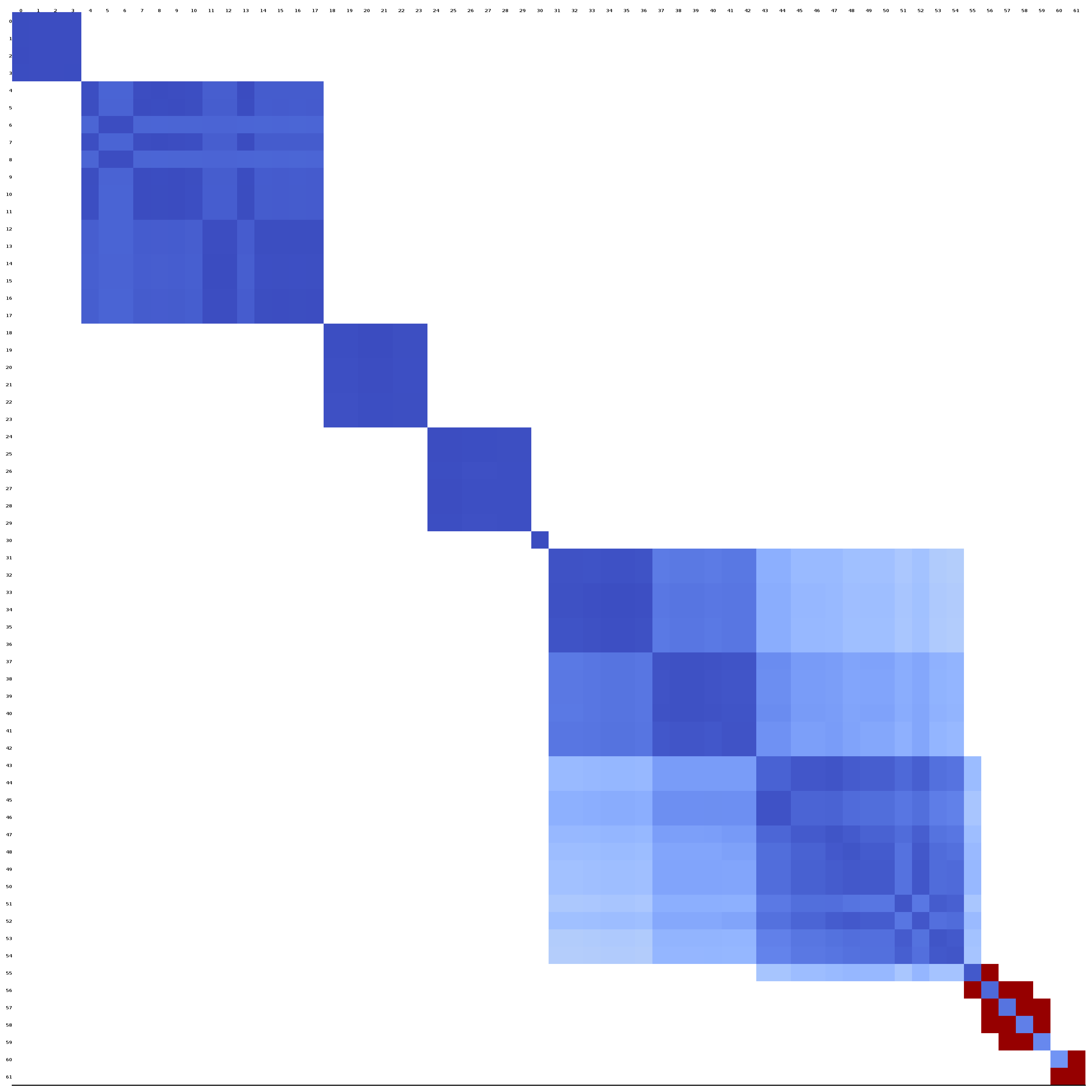}
\label{fig:sub35}}

\subfigure[Zoomed DM for  $(T_3,T_4)$]{\includegraphics[width=0.25\textwidth]{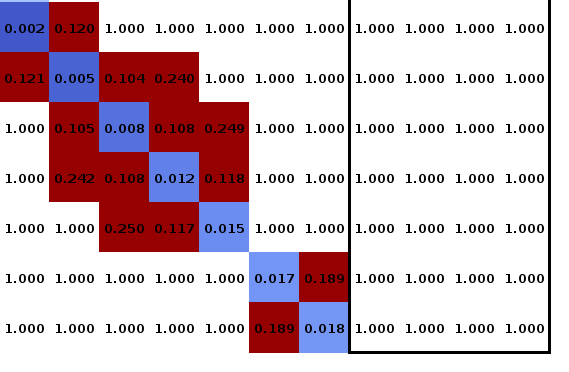}
\label{fig:subz34}}
~
\subfigure[Zoomed DM for  $(T_3,T_5)$ ]{\includegraphics[width=0.16\textwidth]{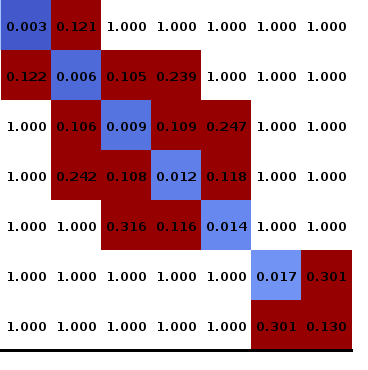}
\label{fig:subz35}}

\caption{Measuring the effect of subsampling using \LMTED. We use a blue-red colormap for the distances ($0$~\protect\includegraphics[height=0.15cm]{Images/legend2.png}~$0.1$). Entries that are discarded due to the refinement are marked with $1.000$.}
\label{fig:subrow3}
\vspace{-0.15in}
\end{figure}

The non-monotonic variation of the distance along a row / column can be due to multiple factors. While the subsampling and smoothing affects the number of critical points, and therefore affects the distance, it is not the only deciding factor. The distance is also affected by (a)~type of critical points inserted / removed, (b)~their function values, and (c)~changes in persistence and pairing. We construct the DMs of the \MTED and \LMTED for the subsampled functions. To highlight the utility of \LMTED, we pick the non-monotonic entries indexed $(3,4), (3,5), (3,6)$ from Figure~\ref{fig:sub2}. The trees are $|T_3|= 62, |T_4|= 66, |T_5|= 62, |T_6|= 66$. We observe that $|T_3|= |T_5|$ but $D_c(T_3,T_5) > D_c(T_3,T_4)$ and $D_c(T_3,T_5) > D_c(T_3,T_6)$ even though  $|T_3|\ne |T_4|, |T_3|\ne |T_6|$. Thus, size cannot explain the non-monotonicity. Also, we notice that $T_3$ and $T_5$ are structurally similar, all edits are relabels and there is negligible difference in the function values of the critical points too. The DMs (Figures~\ref{fig:sub34},\ref{fig:sub35}) show small changes in the pattern, but the values are similar. The bottom-right portions of the DMs along with the values are shown in Figures~\ref{fig:subz34}, and~\ref{fig:subz35}. The diagonal entries in left portion of Figure~\ref{fig:subz34} related to $(3,4)$ shows a gradual increase, while in case of  Figure~\ref{fig:subz35} related to $(3,5)$ we observe an upward spike in the last entry. The corresponding entries for \MTED in both cases change gradually, even though for $(3,5)$ the increase is higher. \LMTED uses truncated persistence for all subtrees and effect of change in persistence pairings is seen only in the global comparison, causing a jump. So, the change in distance means that the subsampling has caused a change in persistence pairing when we go from resolution $4$ to $5$ and $5$ to $6$ but no such change when we go from $3$ to $4$. Observation of the pairings confirms this. We also saw that the pairing changes in $4$ to $5$ was reversed from $5$ to $6$, thus resulting in a lower value in the entry $(3,6)$.

\myparagraph{Fine-grained analysing using \LMTED. } 
\LMTED can be used in conjunction with \MTED to quantify the changes caused by subsampling (or smoothing). This is achieved by computing \MTED across all resolutions and checking if the variation is monotonic. If yes, then the subsampling is likely to have caused changes only in terms of (a)~the number of critical points, (b)~the function values of the critical points, and (c)~persistence of critical points. If the variation is non-monotonic with a jump in the last entries of corresponding \LMTED, then irrespective of other factors, there are changes in persistence pairing resulting in changed matching costs and large changes in distance. Due to the use of truncated persistence in \LMTED, we can detect such changes as jumps in distances. While both \MTED and \LMTED may be unstable, we observe in practice that they are more discriminative than bottleneck and Wasserstein distances.
\begin{figure}
   \centering
   \includegraphics[width=0.45\textwidth]{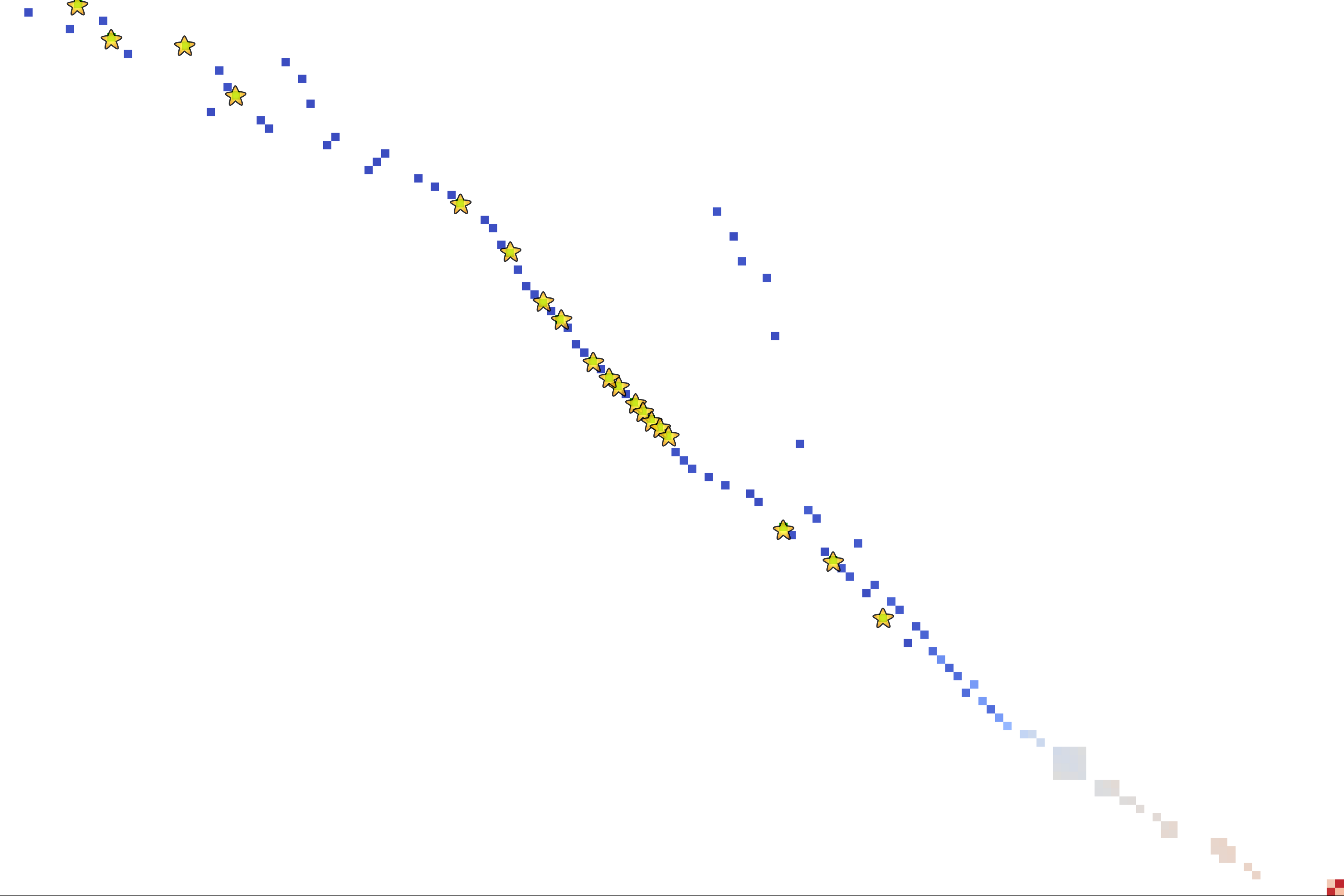}
    \caption{Topological effects of compression. Yellow stars in the DM for the subtree pair $T_{0.5},T_1$ correspond to regions that remain unchanged. Distances in the DM are shown using a blue-red colormap ($0$~\protect\includegraphics[height=0.15cm]{Images/legend2.png}~$0.4$).}
    \label{fig:topcompDM}
    \vspace{-0.15in}
\end{figure}

\myparagraph{Effect of topologically controlled lossy compression. }
Soler et al.~\cite{Soler2018a} describe a method to compress scalar fields that guarantees topology preservation. The method ensures that the bottleneck distance between the persistence diagrams of the compressed and uncompressed field is less than a user specified threshold. Naturally, the method does not consider spatial or hierarchical structure since it is restricted to the persistence diagrams. We present here a fine-grained analysis of the effects of compression using \LMTED.  Soler et al. employ a topological compression followed by zfp. In our experiments, we use only the former. We begin by computing merge trees for both the compressed ($T_c$) and uncompressed data ($T_u$). Since the two scalar fields are defined over a common domain, we select pairs of subtrees of $T_c$ and $T_u$ that correspond to the same region in the domain and order them based on region size. We compute \LMTED between these subtree pairs and note that as we move up the tree hierarchy, the distance remains $0.0$ for some pairs. The largest among the pairs represent regions that remain unchanged post compression. Other \LMTED values follow a staircase pattern, staying level for a few pairs followed by a jump in value. The jump indicates that compression has caused a change in the corresponding subtree. Thus we may identify and isolate regions where compression has no effect in terms of the function value followed by regions that are affected, and traversing the hierarchy of the merge tree lends itself to a multi-scale analysis of the effects of topological compression.

\begin{figure}
    \centering
    \includegraphics[width=0.45\textwidth]{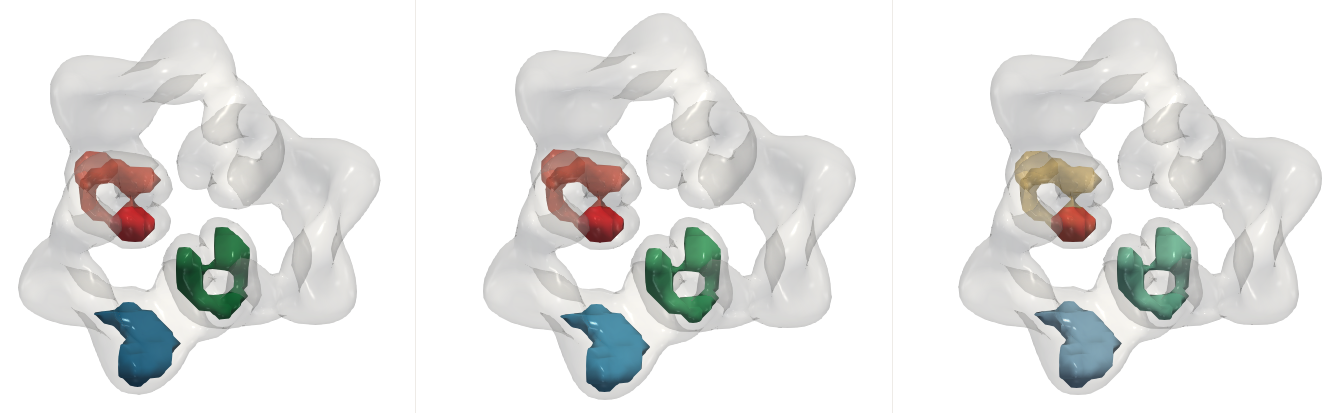}
    \caption{CryoEM image of AMP-Activated Protein Kinase (EMDB-1897)  using different compression thresholds - $0.5\%, 1\%,$ and $2\%$. The region in red is the largest region that remains unaffected, the region in light red is the largest region that remains unaffected for compression threshold of $1\%$, and orange corresponds to $2\%$. Regions in shades of green are symmetric to the regions in light red and orange but are affected by the compression. The regions in shades of blue are also affected by compression. The entire protein is rendered grey and transparent for context.}
    \label{fig:comp}
    \vspace{-0.20in}
\end{figure}

We show results of our analysis applied on AMP-Activated Protein Kinase (EMDB-1897). We apply topological compression using compression thresholds $0.5\%,1\%,2\%$, and compute merge trees $T_{0.5},T_1,T_2$ using TTK~\cite{tierny2018}. To reduce the tree sizes in the experiment, we consider $T_{0.5}$ as the baseline uncompressed data. We choose regions with $100\%$ overlap and compute \LMTED. In Figure~\ref{fig:comp}, we highlight region(s)  that remain unchanged for various thresholds of compression at multiple scales together with a region that is affected due to compression. Figure~\ref{fig:topcompDM} shows the DM for the subtree pair $T_{0.5},T_1$, highlighting unchanged regions by a yellow star. We notice that 19 regions remain unchanged between  $T_{0.5}$ and $T_1$, and 3 regions remain unchanged between $T_{0.5}$ and $T_2$. 
A threshold on \LMTED may be used to highlight regions that are either affected or remain unaffected for various compression thresholds.

\subsection{Spatio-temporal exploration and feature tracking}
\begin{figure*}

 \centering
 
 \subfigure{\includegraphics[width=0.18\textwidth]{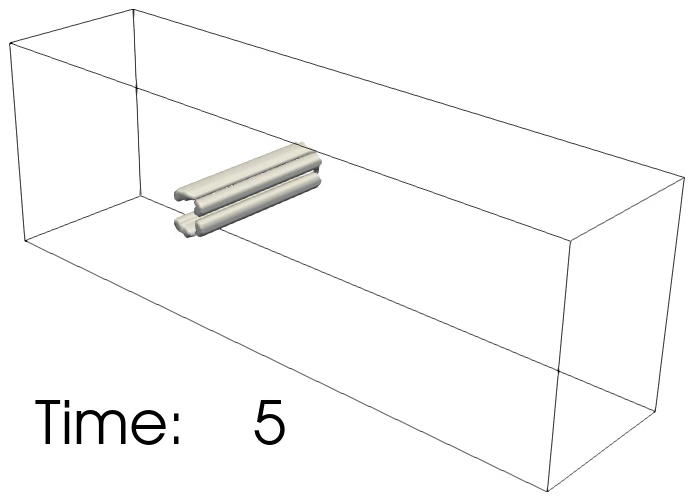}
\label{fig:t5}}
~
\subfigure{\includegraphics[width=0.18\textwidth]{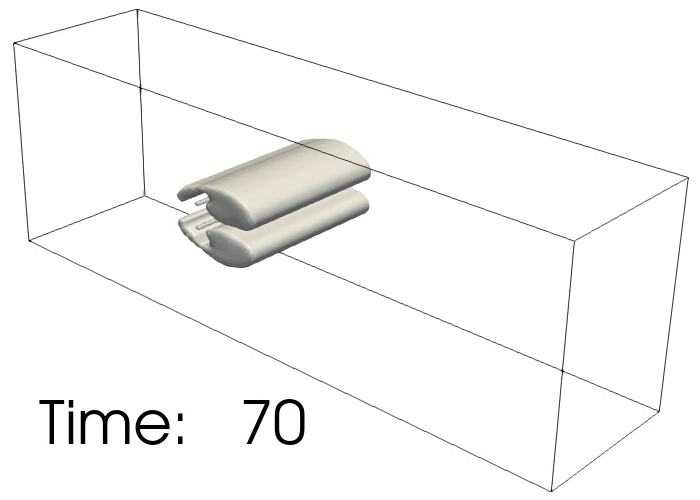}
\label{fig:t70}}
~
\subfigure{\includegraphics[width=0.18\textwidth]{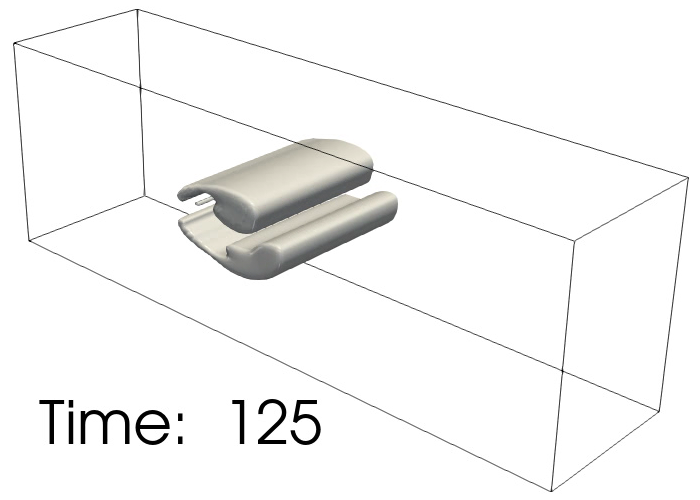}
\label{fig:t125}}
~
\subfigure{\includegraphics[width=0.18\textwidth]{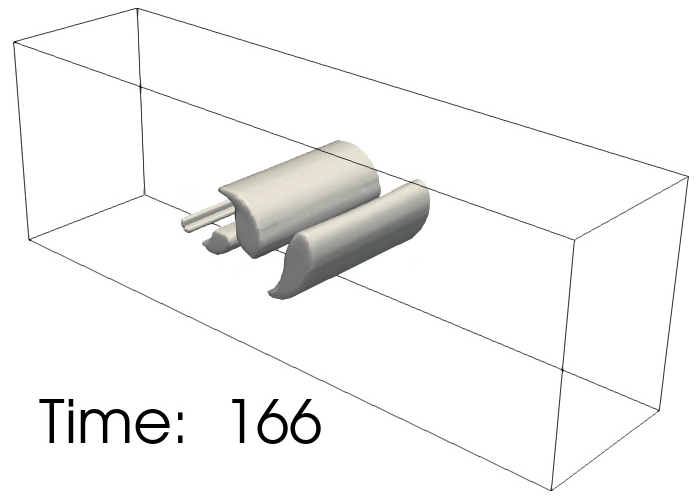}
\label{fig:t166}}
~
\subfigure{\includegraphics[width=0.18\textwidth]{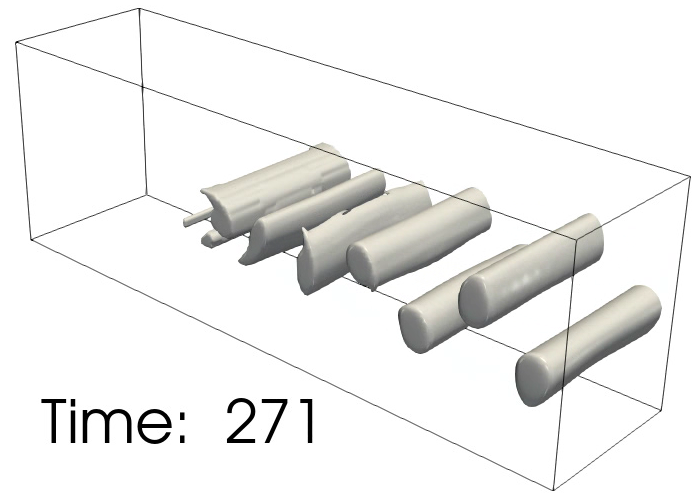}
\label{fig:t271}}
~
\subfigure{\includegraphics[width=0.18\textwidth]{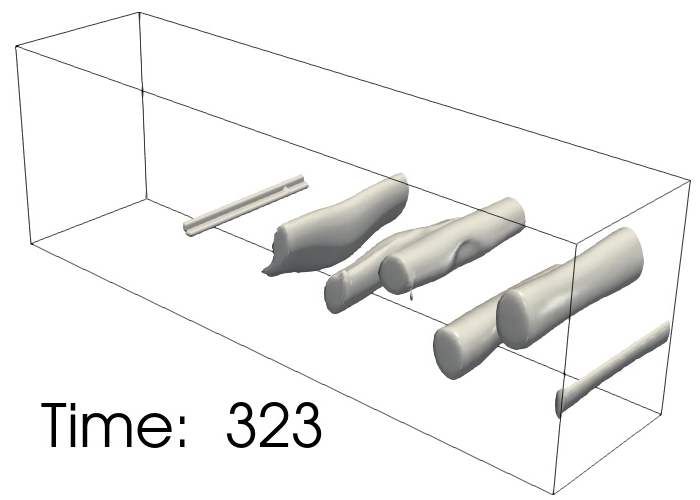}
\label{fig:t323}}
~
\subfigure{\includegraphics[width=0.18\textwidth]{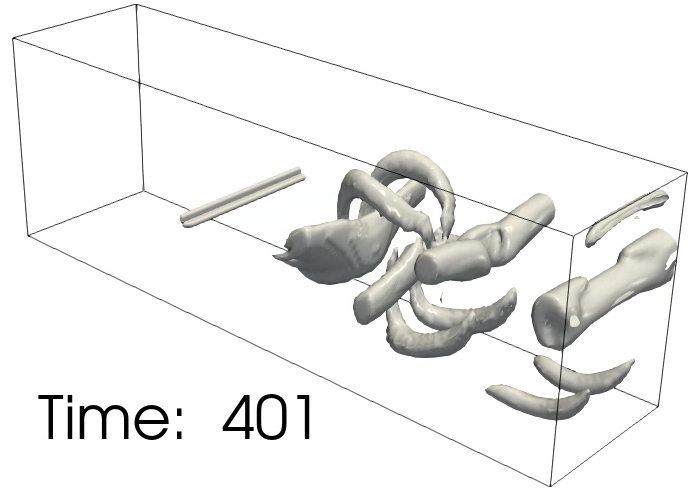}
\label{fig:t401}}
~
\subfigure{\includegraphics[width=0.18\textwidth]{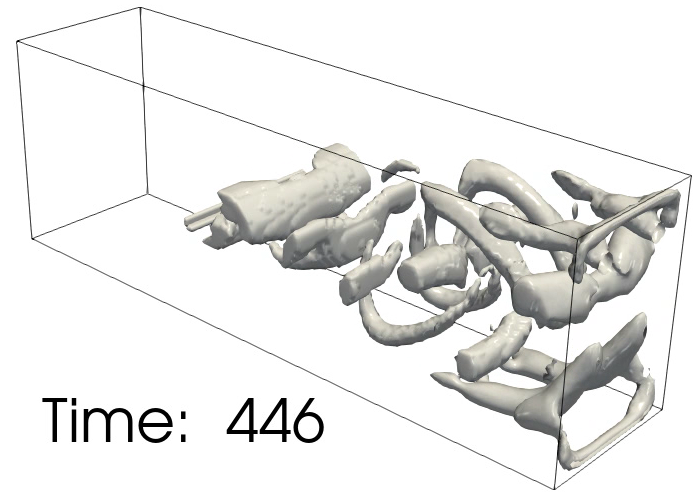}
\label{fig:t446}}
~
\subfigure{\includegraphics[width=0.18\textwidth]{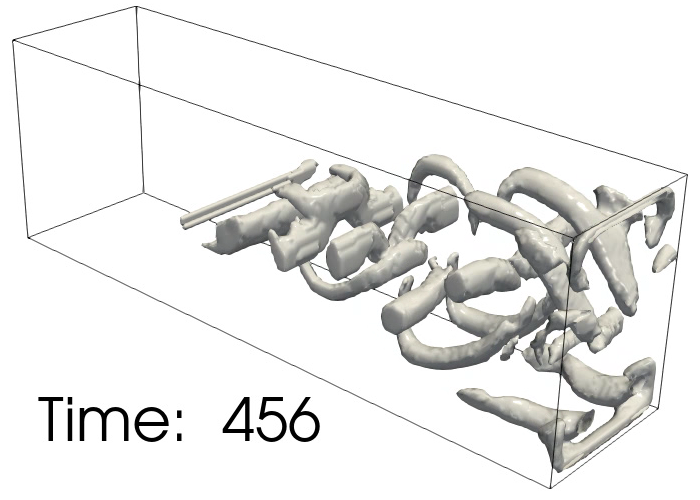}
\label{fig:t456}}
~
\subfigure{\includegraphics[width=0.18\textwidth]{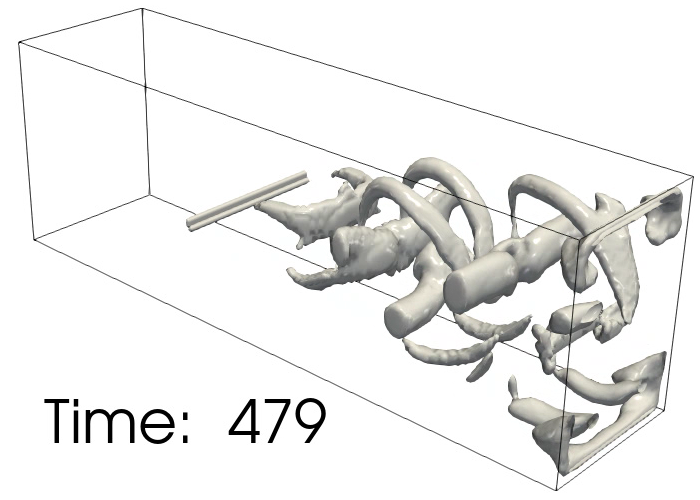}
\label{fig:t479}}

  \caption{Visualizing the top $k$ tracks in the 3D von K\'arm\'an vortex street data. The tracks are generated based on \LMTED and spatial overlaps, and sorted based on the weights of the tracks. The top tracks capture the temporal evolution of a set of primary and secondary vortices.}

  \label{fig:total}

\end{figure*}
\begin{figure*}

 \centering
 
 \subfigure[Query, a primary and a secondary vortex]{\includegraphics[width=0.18\textwidth]{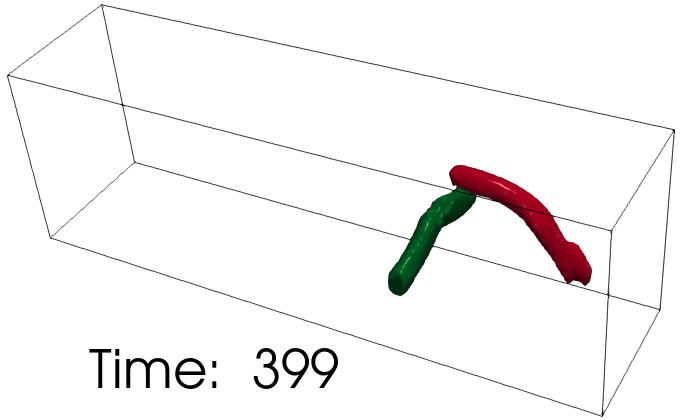}
\label{fig:query}}
~
\rulesep
\subfigure{\includegraphics[width=0.18\textwidth]{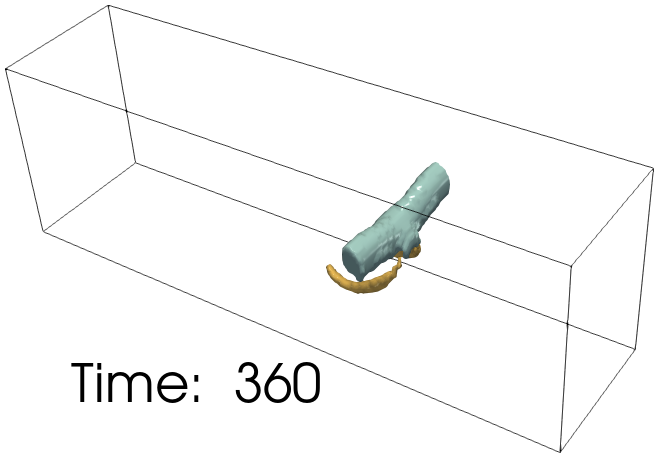}
\label{fig:360}}
~
\subfigure{\includegraphics[width=0.18\textwidth]{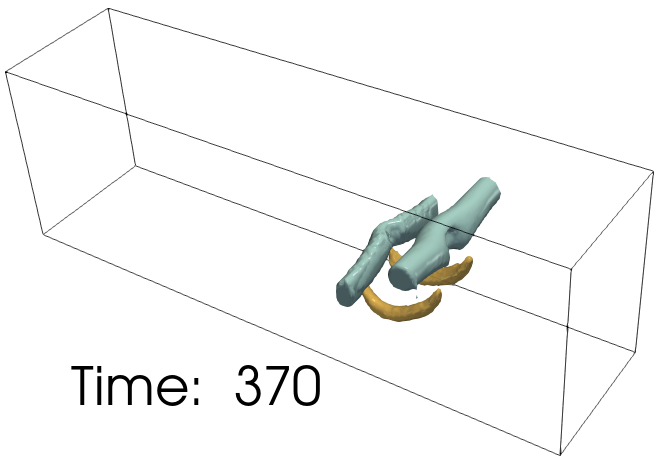}
\label{fig:370}}
~
\subfigure{\includegraphics[width=0.18\textwidth]{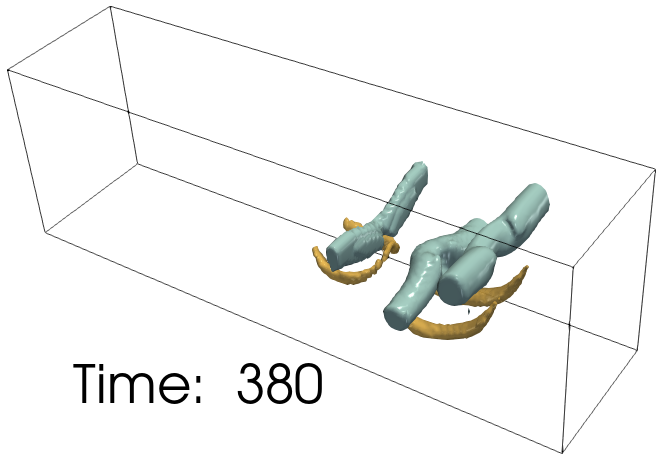}
\label{fig:380}}
~
\subfigure{\includegraphics[width=0.18\textwidth]{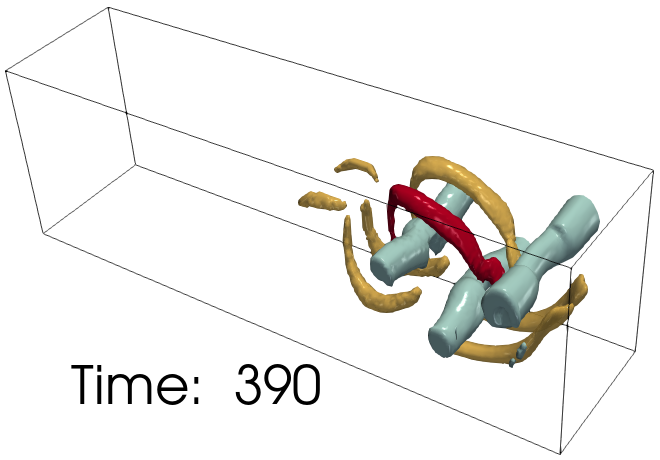}
\label{fig:390}}
~
\subfigure[Symmetric vortices]{\includegraphics[width=0.18\textwidth]{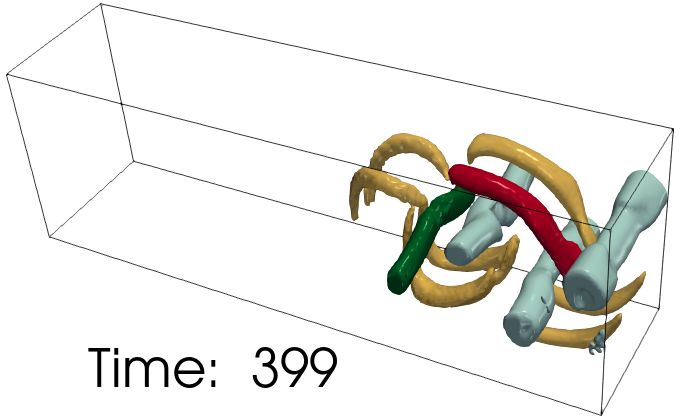}
\label{fig:symm}}
~
\rulesep
\subfigure{\includegraphics[width=0.18\textwidth]{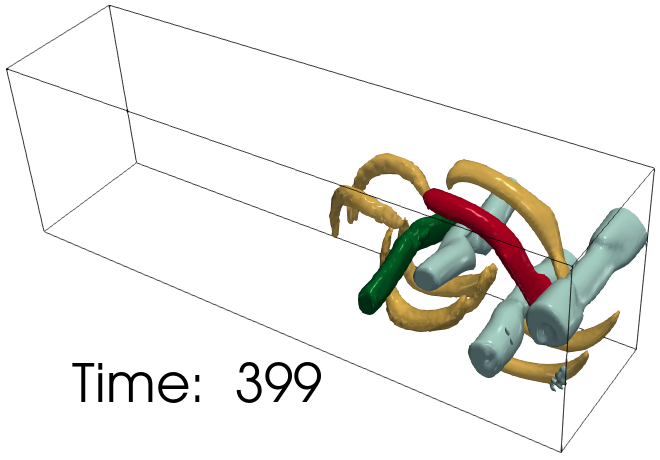}
\label{fig:399}}
~
\subfigure{\includegraphics[width=0.18\textwidth]{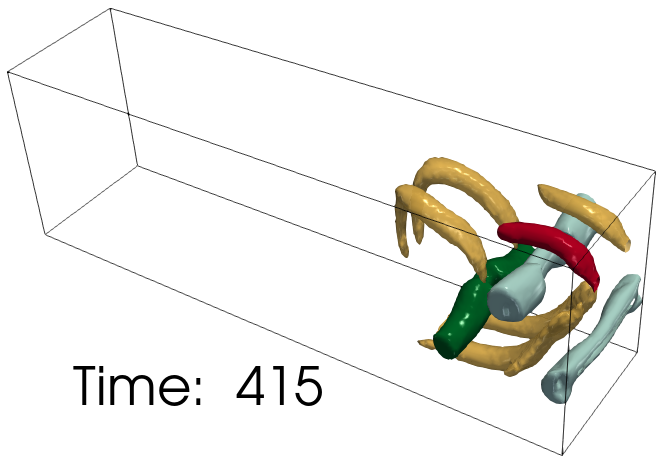}
\label{fig:415}}
~
\subfigure{\includegraphics[width=0.18\textwidth]{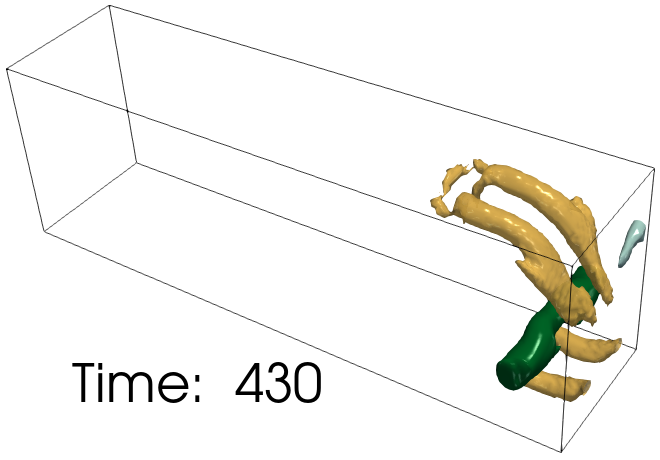}
\label{fig:430}}
~
\subfigure{\includegraphics[width=0.18\textwidth]{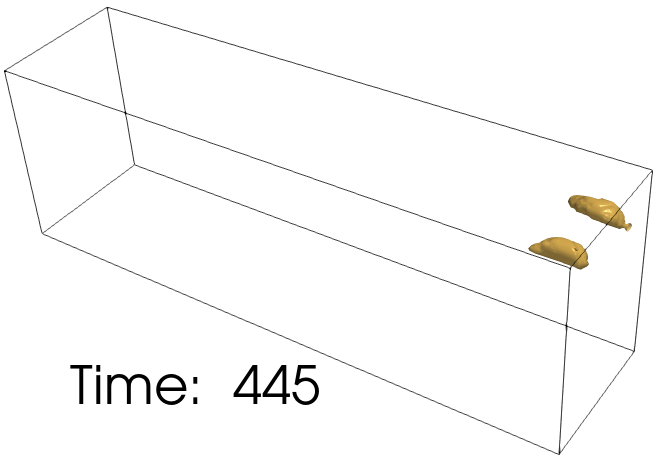}
\label{fig:445}}

  \caption{Tracking query regions in the 3D von K\'arm\'an vortex street data. A query containing a primary (green) and secondary (red) vortex in time step~399 is selected~\subref{fig:query}. \LMTED is used to compute regions (light green and orange) in the same symmetry class as the query~\subref{fig:symm}. All regions in the symmetry class are tracked backward in time (Top row), and forward in time (Bottom row).}

  \label{fig:fin}
\vspace{-0.15in}
\end{figure*}
We demonstrate an application of \LMTED to time-varying scalar fields, in particular for identifying and tracking features across time. We consider two scenarios -- identify and track all features to provide an overview and an interactive query-driven mode for feature tracking.

In order to identify and track all interesting topological features, we begin by computing a sequence of \LMTED DMs between consecutive timesteps.  We apply the refinements described in Section~\ref{sec:refinementoptimization} and compute spatial overlaps between regions that correspond to the reduced set of subtree pairs. We construct a track graph whose nodes represent each region and insert an edge between two nodes in consecutive timesteps if there is a significant overlap between the corresponding regions. Long paths in the track graph correspond to long-lived features. We visualize all long-lived features and their evolution over time including birth, death, split, and merge events. The individual tracks are also used as a starting point for further analysis.
An alternative approach is to allow the user to specify one or many features within a particular timestep. We compute regions that are symmetric to the given feature, compute tracks for each of these regions, and visualize the tracks.

We demonstrate both scenarios using a 3D B\'enard-von K\'arm\'an vortex street dataset. The Okubo-Weiss criterion, indicative of high vorticity regions, is sampled on a regular grid~\cite{saikia2017}. The scalar field is available on $192 \times 64 \times 48$ grid with $508$ timesteps. We compute merge trees for all  timesteps and simplify them using a small persistence threshold of $0.8\%$ to remove noise. We compute \LMTED on the simplified merge trees after applying the appropriate refinement steps mentioned in Section~\ref{sec:refinementoptimization}. The weight of an edge in the track graph is set equal to the spatial overlap (volume of overlap normalized by the volume of union) between the corresponding regions. Overlaps below a $2\%$ threshold are considered negligible and not included into the track graph.

We process the track graph to enumerate top tracks ordered by either the length of the track or the sum of weights (high to low). Further, short tracks (length $<10$) and tracks whose sum of weights is low ($<3.0$) are removed from consideration. We observe that the first track is a thin region close to the cylinder obstruction, which remains almost stationary. Other tracks that appear at the top of the list include the primary and secondary vortices as identified by~\cite{saikia2017}, see Figure~\ref{fig:total}. These vortices are represented as isosurfaces (isovalue 0.1). 

In the second scenario, we use a primary and secondary vortex from timestep $399$ as a query feature, see Figure~\ref{fig:fin}. First, we compute symmetric regions within the same timestep in order to highlight other primary and secondary vortices. Next, we compute tracks that contain the query regions and visualize them. We observe that in the first step \LMTED can discriminate between the primary and secondary vortices and, next, it helps efficiently track the features (vortices) over time. This demonstrates the utility of \LMTED in the exploration of time-varying data. 

The accompanying video in supplementary material shows (a) the top tracks corresponding to the primary and secondary vortices, (b) query based exploration. 

There are a few exceptional situations where \LMTED is unable to discriminate between primary and secondary vortices. This happens when, say, the chosen vortex is a secondary vortex, corresponds to a leaf node in the merge tree, and matches with a leaf node that corresponds to a primary vortex. Further spatial overlap tests are necessary to identify that the two regions do not correspond to each other.
To summarize, \LMTED supports the generation of a good overview visualization and serves as a starting point for feature detection and tracking. Subsequent interaction and visualization tasks are often necessary and these tasks may closely depend on application specific requirements.

%% file: section7.tex
\section{Conclusions}
We described a local comparison measure (\LMTED) between two scalar fields by comparing subtrees of their merge trees. The comparison measure supports local and fine-grained analysis and visualization of similarities and differences between two scalar fields. The measure satisfies metric properties and can be efficiently computed. We demonstrate its practical utility via applications to feature tracking, study of topology controlled compression, and symmetry identification. In future work, we plan to develop a comparative visualization framework based on the \MTED and \LMTED that may be applied to time-varying and ensemble data. 

%% file: localsimilarity.bbl
 \newcommand{\noop}[1]{}
\begin{thebibliography}{10}
\providecommand{\url}[1]{#1}
\csname url@samestyle\endcsname
\providecommand{\newblock}{\relax}
\providecommand{\bibinfo}[2]{#2}
\providecommand{\BIBentrySTDinterwordspacing}{\spaceskip=0pt\relax}
\providecommand{\BIBentryALTinterwordstretchfactor}{4}
\providecommand{\BIBentryALTinterwordspacing}{\spaceskip=\fontdimen2\font plus
\BIBentryALTinterwordstretchfactor\fontdimen3\font minus
  \fontdimen4\font\relax}
\providecommand{\BIBforeignlanguage}[2]{{%
\expandafter\ifx\csname l@#1\endcsname\relax
\typeout{** WARNING: IEEEtran.bst: No hyphenation pattern has been}%
\typeout{** loaded for the language `#1'. Using the pattern for}%
\typeout{** the default language instead.}%
\else
\language=\csname l@#1\endcsname
\fi
#2}}
\providecommand{\BIBdecl}{\relax}
\BIBdecl

\bibitem{cohen2007}
D.~Cohen-Steiner, H.~Edelsbrunner, and J.~Harer, ``Stability of persistence
  diagrams,'' \emph{Discrete and Computational Geometry}, vol.~37, no.~1, pp.
  103--120, 2007.

\bibitem{di2012}
B.~Di~Fabio and C.~Landi, ``Stability of {R}eeb graphs of closed curves,''
  \emph{Electronic Notes in Theoretical Computer Science}, vol. 283, pp.
  71--76, 2012.

\bibitem{morozov2013}
D.~Morozov, K.~Beketayev, and G.~Weber, ``Interleaving distance between merge
  trees,'' \emph{Discrete and Computational Geometry}, vol.~49, no.~52, pp.
  22--45, 2013.

\bibitem{bauer2014}
U.~Bauer, X.~Ge, and Y.~Wang, ``Measuring distance between {R}eeb graphs,'' in
  \emph{Proceedings of the 13th annual Symposium on Computational
  Geometry}.\hskip 1em plus 0.5em minus 0.4em\relax ACM, 2014, pp. 464--474.

\bibitem{beketayev2014}
K.~Beketayev, D.~Yeliussizov, D.~Morozov, G.~H. Weber, and B.~Hamann,
  ``Measuring the distance between merge trees,'' in \emph{TopoinVis
  III}.\hskip 1em plus 0.5em minus 0.4em\relax Springer, 2014, pp. 151--165.

\bibitem{Saikia2014}
H.~Saikia, H.~P. Seidel, and T.~Weinkauf, ``{Extended branch decomposition
  graphs: Structural comparison of scalar data},'' \emph{Computer Graphics
  Forum}, vol.~33, no.~3, pp. 41--50, 2014.

\bibitem{saikia2015}
H.~Saikia, H.-P. Seidel, and T.~Weinkauf, ``Fast similarity search in scalar
  fields using merging histograms,'' in \emph{TopoInVis 2015}, 2015, pp. 1--14.

\bibitem{dey2015}
T.~Dey, D.~Shi, and Y.~Wang, ``{Comparing Graphs via Persistence Distortion},''
  \emph{31st International Symposium on Computational Geometry (SoCG 2015)},
  pp. 491--506, 2015.

\bibitem{narayanan2015}
V.~Narayanan, D.~M. Thomas, and V.~Natarajan, ``Distance between extremum
  graphs,'' in \emph{PacificVis}, 2015, pp. 263--270.

\bibitem{di2016}
B.~Di~Fabio and C.~Landi, ``The edit distance for {R}eeb graphs of surfaces,''
  \emph{Discrete and Computational Geometry}, vol.~55, no.~2, pp. 423--461,
  2016.

\bibitem{saikia2017}
H.~Saikia and T.~Weinkauf, ``Global feature tracking and similarity estimation
  in time-dependent scalar fields,'' \emph{Computer Graphics Forum}, vol.~36,
  no.~3, pp. 1--11, 2017.

\bibitem{Sridh2017}
R.~Sridharamurthy, A.~Kamakshidasan, and V.~Natarajan, ``Edit distances for
  comparing merge trees,'' in \emph{IEEE SciVis Posters}, 2017.

\bibitem{Sridh2020}
R.~Sridharamurthy, T.~B. Masood, A.~Kamakshidasan, and V.~Natarajan, ``Edit
  distance between merge trees,'' \emph{{IEEE} Transactions on Visualization
  and Computer Graphics}, vol.~26, no.~3, pp. 1518--1531, 2020.

\bibitem{tao2018}
J.~Tao, M.~Imre, C.~Wang, N.~V. Chawla, H.~Guo, G.~Sever, and S.~H. Kim,
  ``Exploring time-varying multivariate volume data using matrix of isosurface
  similarity maps,'' \emph{IEEE transactions on visualization and computer
  graphics}, vol.~25, no.~1, pp. 1236--1245, 2018.

\bibitem{Bruckner2010}
S.~Bruckner and T.~M{\"{o}}ller, ``{Isosurface similarity maps},''
  \emph{Computer Graphics Forum}, vol.~29, no.~3, pp. 773--782, 2010.

\bibitem{lukasczyk2017}
J.~Lukasczyk, G.~Weber, R.~Maciejewski, C.~Garth, and H.~Leitte, ``Nested
  tracking graphs,'' \emph{Computer Graphics Forum}, vol.~36, no.~3, pp.
  12--22, 2017.

\bibitem{lukasczyk2019}
J.~Lukasczyk, C.~Garth, G.~H. Weber, T.~Biedert, R.~Maciejewski, and H.~Leitte,
  ``Dynamic nested tracking graphs,'' \emph{IEEE Transactions on Visualization
  and Computer Graphics}, vol.~26, no.~1, pp. 249--258, 2019.

\bibitem{thomas2011}
D.~M. Thomas and V.~Natarajan, ``Symmetry in scalar field topology,''
  \emph{IEEE Transactions on Visualization and Computer Graphics}, vol.~17,
  no.~12, pp. 2035--2044, 2011.

\bibitem{thomas2013}
------, ``Detecting symmetry in scalar fields using augmented extremum
  graphs,'' \emph{IEEE Transactions on Visualization and Computer Graphics},
  vol.~19, no.~12, pp. 2663--2672, 2013.

\bibitem{thomas2014}
------, ``Multiscale symmetry detection in scalar fields by clustering
  contours,'' \emph{IEEE Transactions on Visualization and Computer Graphics},
  vol.~20, no.~12, pp. 2427--2436, 2014.

\bibitem{carr2003}
H.~Carr, J.~Snoeyink, and U.~Axen, ``Computing contour trees in all
  dimensions,'' \emph{Computational Geometry}, vol.~24, no.~2, pp. 75--94,
  2003.

\bibitem{edelsbrunner2000}
H.~Edelsbrunner, D.~Letscher, and A.~Zomorodian, ``Topological persistence and
  simplification,'' in \emph{{F}oundations of {C}omputer {S}cience}.\hskip 1em
  plus 0.5em minus 0.4em\relax IEEE, 2000, pp. 454--463.

\bibitem{acharya2015parallel}
A.~Acharya and V.~Natarajan, ``A parallel and memory efficient algorithm for
  constructing the contour tree,'' in \emph{PacificVis}, 2015, pp. 271--278.

\bibitem{Morozov2013dist}
D.~Morozov and G.~Weber, ``Distributed merge trees,'' in \emph{Proc. ACM
  SIGPLAN Symposium on Principles and Practice of Parallel Programming}, ser.
  PPoPP '13, 2013, pp. 93--102.

\bibitem{Gueunet2017}
C.~Gueunet, P.~Fortin, J.~Jomier, and J.~Tierny, ``Task-based augmented merge
  trees with fibonacci heaps,'' in \emph{2017 IEEE 7th Symposium on Large Data
  Analysis and Visualization (LDAV)}.\hskip 1em plus 0.5em minus 0.4em\relax
  IEEE, 2017, pp. 6--15.

\bibitem{Bille2005}
P.~Bille, ``{A survey on tree edit distance and related problems},''
  \emph{Theoretical Computer Science}, vol. 337, no. 1-3, pp. 217--239, 2005.

\bibitem{Zhang1996}
K.~Zhang, ``{A Constrained Edit Distance Between Unordered Labeled Trees},''
  \emph{Algorithmica}, vol.~15, pp. 205--222, 1996.

\bibitem{Zhang1992}
K.~Zhang, R.~Statman, and D.~Shasha, ``{On the editing distance between
  unordered labeled trees},'' \emph{Information Processing Letters}, vol.~42,
  no.~3, pp. 133--139, 1992.

\bibitem{EMDB2021}
``Protein data bank in {E}urope,'' \url{https://www.ebi.ac.uk/pdbe/emdb/},
  2021, accessed: 20-03-2021.

\bibitem{behrisch2016}
M.~Behrisch, B.~Bach, N.~Henry~Riche, T.~Schreck, and J.-D. Fekete, ``Matrix
  reordering methods for table and network visualization,'' \emph{Computer
  Graphics Forum}, vol.~35, no.~3, pp. 693--716, 2016.

\bibitem{Soler2018a}
M.~Soler, M.~Plainchault, B.~Conche, and J.~Tierny, ``{Topologically Controlled
  Lossy Compression},'' \emph{IEEE Pacific Visualization Symposium}, pp.
  46--55, 2018.

\bibitem{tierny2018}
J.~Tierny, G.~Favelier, J.~A. Levine, C.~Gueunet, and M.~Michaux, ``The
  {T}opology {T}ool{K}it,'' \emph{IEEE Transactions on Visualization and
  Computer Graphics}, vol.~24, no.~1, pp. 832--842, 2018,
  \url{https://topology-tool-kit.github.io/}.

\end{thebibliography}
